\documentclass[a4paper,11pt]{article}
\pdfoutput=1 

\usepackage{jheppub} 

\usepackage[T1]{fontenc} 
\usepackage[UKenglish]{babel}

\usepackage{xcolor}
\usepackage{amsmath}
\usepackage{tensor}
\usepackage[utf8]{inputenc}

\title{Consistent Searches for SMEFT Effects in Non-Resonant Dijet Events}
\preprint{\begin{flushright}FERMILAB-PUB-17-507-T\\MITP/17-082\\ZU-TH-34/17\end{flushright}}
\author{Stefan Alte${}^a$,}
\author{Matthias K\"onig${}^{a,b}$,}
\author[a]{and William Shepherd}

\emailAdd{stalte@uni-mainz.de}
\emailAdd{matthias.koenig@uzh.ch}
\emailAdd{shepherd@uni-mainz.de}

\affiliation[a]{PRISMA Cluster of Excellence \& Mainz Institute of Theoretical Physics, \\ Johannes Gutenberg-Universit\"at Mainz, 55099 Mainz, Germany}
\affiliation[b]{Physik-Institut, Universit\"at Z\"urich, CH-8057, Switzerland}

\abstract{
We investigate the bounds which can be placed on generic new-physics contributions to dijet production at the LHC using the framework of the Standard Model Effective Field Theory, deriving  the first consistently-treated EFT bounds from non-resonant high-energy data. We recast an analysis searching for quark compositeness, equivalent to treating the SM with one higher-dimensional operator as a complete UV model. In order to reach consistent, model-independent EFT conclusions, it is necessary to truncate the EFT effects consistently at order $1/\Lambda^2$ and to include the possibility of multiple operators simultaneously contributing to the observables, neither of which has been done in previous searches of this nature. Furthermore, it is important to give consistent error estimates for the theoretical predictions of the signal model, particularly in the region of phase space where the probed energy is approaching the cutoff scale of the EFT. There are two linear combinations of operators which contribute to dijet production in the SMEFT with distinct angular behavior; we identify those linear combinations and determine the ability of LHC searches to constrain them simultaneously. Consistently treating the EFT generically leads to weakened bounds on new-physics parameters. These constraints will be a useful input to future global analyses in the SMEFT framework, and the techniques used here to consistently search for EFT effects are directly applicable to other off-resonance signals.
}

\begin{document}

\maketitle

\section{Introduction}

The LHC continues to produce record-setting amounts of data, and corresponding record-setting limits on the various concrete models that have been proposed for physics beyond the Standard Model (BSM). In particular, colored new particles are now generically constrained to be in the multi-TeV range at the lightest. At the same time, a huge amount of data at lower partonic center-of-mass energies is becoming available for high-precision studies of the dynamics of Standard Model (SM) particles. Given this combination, it is only natural that the approach of using Effective Field Theory (EFT) techniques to parameterize the effects of heavy new physics (NP) on the lower-energy dynamics has grown rapidly.

There exist two consistent EFT approaches that can describe such heavy NP well below its characteristic mass scale. The most general such theory is known as Higgs Effective Field Theory (HEFT), which treats electroweak (EW) symmetry akin to the way that chiral symmetry is treated in the chiral perturbation theory of low-energy QCD. This means in particular that the full SM gauge group is not manifest in the operators of the HEFT. The physical reasoning for such an omission is that we do not yet have very strong constraints on the Higgs-like scalar discovered and under scrutiny at the LHC; if it is not fully embedded in an $SU(2)_L$ doublet then it is impossible to consistently insist upon that symmetry in the infrared. The Standard Model EFT (SMEFT), in contrast, assumes that the LHC discovery was in fact of the Higgs boson, i.e. that the scalar we found is the remnant of the scalar doublet responsible for breaking EW symmetry in the SM, and that NP at some higher scale does not mix with the Higgs directly, but instead only alters its interactions (and those of other particles in the SM) \cite{Brivio:2017vri}. The SMEFT is formally a subset of the HEFT, as it is possible to construct a limit of the HEFT parameters where EW symmetry is equally manifest.

Analyses of the full set of precision EW and Higgs decay data are now available at tree level in multiple different bases and making various assumptions about the theoretical precision of the predicted deviations \cite{Ellis:2014dva,Falkowski:2014tna,Falkowski:2015jaa,Berthier:2015gja,Butter:2016cvz,Berthier:2016tkq}. It is important to understand, however, that these theoretical errors are definitely not negligible given the exquisite precision of some of the measurements that have been made at the $Z$ pole, and that will be made in the future at the Higgs pole as well; loop corrections in relatively large SM couplings are generically percent-level corrections to the leading NP effect, and higher-order EFT effects from either the squared dimension-six EFT amplitude or unknown dimension-eight operators interfering with the SM amplitude are suppressed by $v^2/\Lambda^2\sim\%$ as well.

Attempts to better understand and improve the precision of these predictions have also begun in earnest \cite{Hartmann:2015aia,Hartmann:2015oia,Gauld:2015lmb,Gauld:2016kuu,Freitas:2016iwx,Hartmann:2016pil}, and have highlighted in particular the sensitivity to new operators introduced at loop level. While a tree-level, narrow-width treatment of EW and Higgs data can neglect many operators, most notably all but one of the four-fermion operators which make up nearly half of the basis of flavor-blind and CP-conserving SMEFT operators, a loop-level treatment reintroduces sensitivity to a great number of those operators which did not contribute at leading order. This reintroduces flat directions for all the EW and Higgs data at loop level, and thus prevents us from realistically claiming constraints better than at the few-percent level based on precision observables without independent constraints on these new operators from other data.

Unfortunately, treating EFT contributions without a fixed momentum transfer arising from the measurement being made at a pole is much more subtle, as the EFT corrections grow with energy. One crucial step has already become commonplace in the literature treating these off-shell EFT effects, which is to explicitly discard any data with partonic center-of-mass energy greater than the cutoff scale \cite{Englert:2014cva,Contino:2016jqw,Farina:2016rws,Alioli:2017jdo}. This is of course necessary by construction; any physics beyond the EFT cutoff scale is explicitly not well-described by the EFT. However, it is very common practice to treat the EFT in a power-counting-inconsistent way, by keeping the effect of the EFT amplitude squared. This keeps only one term of many that arise at the same order in the inverse EFT cutoff scale, yields different results for different bases of dimension-six operators, and generically artificially enhances the sensitivity of the search. Also, it is standard practice in these searches to ignore the fact that the signal cross section is not arbitrarily well known, but instead has appreciable uncertainties. This is particularly problematic at higher energies, as the uncertainties from higher order EFT effects grow faster with energy than the signal does.  By judiciously discarding data even from scales approaching the cutoff the errors arising from higher-dimensional effects can be reduced, but they still must be considered as additional errors in the analysis if the result is to be consistently usable in NP interpretations. Finally, these searches are generically done assuming that only one or few operators are present in the EFT, which also artificially strengthens bounds as it limits the possibility of cancellations between different operators; it has been shown \cite{Jiang:2016czg} that it is generically not possible to generate only one operator in the SMEFT from UV-complete models.

Here we improve the treatment of all of these problematic features for searches in the tails of distributions. We focus in this article on dijet searches at the LHC, where models with some of the structure of the SMEFT have long been searched for in the context of quark compositeness \cite{Eichten:1983hw,Eichten:1984eu}, though the usual failings described above are present in these searches. To yield true EFT constraints, we consistently truncate our signal prediction at leading nontrivial order in the EFT cutoff scale and utilize the EFT amplitude squared as an ansatz for the general size and shape of theoretical errors arising at next order in the power counting. We also explore the full set of operators which contribute to dijet production at the LHC, rather than assuming arbitrarily that only one operator is relevant in constructing the search. While this leads to much weaker bounds than those usually reported, we emphasize that these are the bounds which have treated the EFT in an honest way, including the uncertainties inherent in the EFT ansatz. As such, these are the only bounds which can consistently be used to constrain NP interpretations of the data which do not conform to the stringent UV assumptions which are implicitly made in the quark compositeness search.

The remainder of this article is organized as follows: in the next section we will quickly review the SMEFT and calculate its contributions to dijet production. We will then consider in detail the requirements of a consistent analysis in the EFT, including appropriate truncation of the power series in the cutoff scale and the need for, as well as our specific treatment of, theoretical errors on the signal cross sections in Section~\ref{sec:consistency}. In Section~\ref{sec:single}, we discuss and reproduce a recent normalized quark compositeness search and recast it as a consistent search in the SMEFT, but find no consistent constraints. We therefore consider an unnormalized search in Section~\ref{sec:full} and find that it can constrain the EFT. Furthermore, we propose a search in the dijet mass distributions alone in Section \ref{sec:mjj}. Finally, we will conclude with some thoughts for future study in Section \ref{sec:conc}.

\section{Dijet production in SMEFT}
\label{sec:SMEFT}
In the SMEFT, the SM Lagrangian is systematically supplemented by higher-dimensional operators built out of SM fields which are invariant under $SU(3)_C \times SU(2)_L \times U(1)_Y$. Therefore, the underlying SMEFT Lagrangian $\mathcal{L}_\text{SMEFT}$ has the general form
\begin{equation}
\label{eq:lagrangian}
 \mathcal{L}_\text{SMEFT} = \mathcal{L}_\text{SM} + \mathcal{L}^{(5)} + \mathcal{L}^{(6)}+\mathcal{L}^{(7)}+\mathcal{L}^{(8)} + \dots \,,
\end{equation}
where $\mathcal{L}_\text{SM}$ is the SM Lagrangian (with the SM couplings corrected at order $v^2/\Lambda^2$) and $\mathcal{L}^{(i)}$ with $4 < i$ denotes the Lagrangian containing higher-dimensional operators of dimension $i$. These terms have the form
\begin{equation}
 \mathcal{L}^{(i)} = \sum_{k=1}^{N_i} \frac{c_k^{(i)}}{\Lambda^{i-4}} Q_k^{(i)},
\end{equation}
where the number of non-redundant operators is $N_i$, the Wilson coefficient is $c_k^{(i)}$, the operator is $Q_k^{(i)}$ and the scale of NP is $\Lambda$. Bases of operators up to and including dimension eight are known \cite{Weinberg:1979sa,Wilczek:1979hc,Buchmuller:1985jz,Grzadkowski:2010es,Abbott:1980zj,Lehman:2014jma,Lehman:2015coa} and methods to determine a basis of operators at higher dimension have been developed \cite{Henning:2015alf,Lehman:2015via,Henning:2015daa,Henning:2017fpj}. The behavior of the dimension-six operators in the Warsaw basis \cite{Grzadkowski:2010es} under renormalization is also known \cite{Jenkins:2013zja,Jenkins:2013wua,Alonso:2013hga}.

In our application of the SMEFT to dijet production, we consider the CP-even and baryon- and lepton-number-conserving operators which are flavor-diagonal. At dimension five only the neutrino Majorana mass operator appears \cite{Weinberg:1979sa,Wilczek:1979hc} (in the absence of our assumptions), so the leading heavy new-physics contributions to dijet production appear at dimension six. We collect the contributing operators in Table \ref{tab:dijet_operators}.
\begin{table}
\renewcommand{\arraystretch}{1.4}
\centering
\begin{tabular}{rc|ccrc|c}
\cline{1-3} \cline{6-7} 
&$Q^{(1)}_{qq}$ & $ \left( \bar{q}_p \gamma_\mu q_r \right) \left( \bar{q}_s \gamma^\mu q_t \right)$ & &
&$Q^{(3)}_{qq}$ & $ \left( \bar{q}_p \gamma_\mu \tau^I q_r \right) \left( \bar{q}_s \gamma^\mu \tau^I q_t \right)$ \\
&$Q_{uu}$       & $ \left( \bar{u}_p \gamma_\mu u_r \right) \left( \bar{u}_s \gamma^\mu u_t \right)$ & &
&$Q_{dd}$       & $ \left( \bar{d}_p \gamma_\mu d_r \right) \left( \bar{d}_s \gamma^\mu d_t \right)$ \\
*&$Q_{ud}^{(1)}$ & $ \left( \bar{u}_p \gamma_\mu u_r \right) \left( \bar{d}_s \gamma^\mu d_t \right)$ & &
&$Q_{ud}^{(8)}$ & $ \left( \bar{u}_p \gamma_\mu T^A u_r \right) \left( \bar{d}_s \gamma^\mu T^A d_t \right)$ \\
*&$Q_{qu}^{(1)}$ & $ \left( \bar{q}_p \gamma_\mu q_r \right) \left( \bar{u}_s \gamma^\mu u_t \right)$ & &
&$Q_{qu}^{(8)}$ & $ \left( \bar{q}_p \gamma_\mu T^A q_r \right) \left( \bar{u}_s \gamma^\mu T^A u_t \right)$ \\
*&$Q_{qd}^{(1)}$ & $ \left( \bar{q}_p \gamma_\mu q_r \right) \left( \bar{d}_s \gamma^\mu d_t \right)$ & &
&$Q_{qd}^{(8)}$ & $ \left( \bar{q}_p \gamma_\mu T^A q_r \right) \left( \bar{d}_s \gamma^\mu T^A d_t \right)$ \\
* & $Q_G$ & $f^{ABC}G_\mu^{A\nu}G_{\nu}^{B\rho}G_{\rho}^{C\mu}$ & &
&  \\
\end{tabular}
\caption{Dimension-six operators built out of SM fields contributing to dijet production in the Warsaw basis~\cite{Grzadkowski:2010es}. These operators conserve baryon number, lepton number, are potentially flavor diagonal, and conserve CP. Operators marked with an asterisk do not interfere with the QCD amplitude and thus enter our prediction only at order $\Lambda^{-4}$.}
\label{tab:dijet_operators}
\end{table}
The left-handed quark doublets are denoted by $q$ and right-handed quark fields are indicated by $u$ for up-type quarks and $d$ for down-type quarks. The generators of $SU(2)_L$ are $\tau^I=\sigma^I/2$, where \mbox{$I\in\{1,2,3\}$} and $\sigma^I$ is a Pauli matrix, and the generators of $SU(3)_C$ are $T^A$ with \mbox{$A\in\{1,\dots 8\}$}. Furthermore, we introduce the generation indices $p,r,s,t$.

The three-gluon operator $Q_G$ could in principle contribute to dijet production at dimension six. However, as is well known, it does not interfere with the leading-order QCD contribution at leading power in the EFT counting \cite{Simmons:1989zs,Simmons:1990dh}, as it couples gluons with a distinct helicity structure to that present in the SM. Note, however, that techniques to discern the effects of this operator at leading order in EFT power counting in three-jet events have been explored \cite{Dixon:1993xd}. A more complete exploration of this helicity-driven phenomenon of non-interference between certain SMEFT operators and the SM tree-level amplitudes can be found in \cite{Azatov:2016sqh}. Note also that the color-singlet four-quark operators composed of bilinears of distinctly-charged quarks cannot interfere with the QCD amplitude; this is due to the need for there to be quarks with identical weak charges and distinct color for a gluon to couple to. These non-interfering operators are marked with an asterisk in Table~\ref{tab:dijet_operators}.

We consider only flavor-symmetric combinations of quarks to compose the currents of these operators, which means that we assume the Wilson coefficients, properly thought of as tensors of rank four in generation space, have the structure $\delta_{pr}\delta_{st}$. This avoids strong constraints from flavor physics, and is a slightly stronger assumption than Minimal Flavor Violation \cite{DAmbrosio:2002vsn}. We note that searches for quark compositeness generically actually insist on an even stronger assumption, that all four indices are identical, but this does not affect the size of the interference term between the EFT and SM contributions, as they are considering only color-singlet operators and the same argument given above regarding distinct weak charges of quarks applies just as well to distinct flavors. However, adopting the requirement that all four indices are identical would lead to nontrivial differences in the interference terms for color-octet operators.

The current LHC searches for quark compositeness, which we will recast to give constraints on the SMEFT contributions to dijet production, analyze the cross section $\sigma$ differential in the angular variable $\chi=\exp\left(|y_1-y_2|\right)$, where $y_1$ and $y_2$ are the pseudorapidities of the two hardest jets in the detector frame \cite{ATLAS:2012pu,Aad:2015eha,ATLAS:2015nsi,Aaboud:2017yvp,Chatrchyan:2012bf,Khachatryan:2014cja,Sirunyan:2017ygf}. The $\chi$ variable is constructed such that the SM contribution is largely independent of $\chi$, whereas the contribution from the four-fermion contact operator considered in these analyses, notably $Q_{qq}^{(1)}$ from Table~\ref{tab:dijet_operators}, is generally largest for small values of $\chi$. This difference in angular behaviors of the signal and background allows various experimental techniques to be used, including the consideration of normalized distributions, where higher-order QCD and experimental jet energy scale uncertainties can be partially removed.  

\begin{figure}
 \centering
 \includegraphics[width=.52\textwidth]{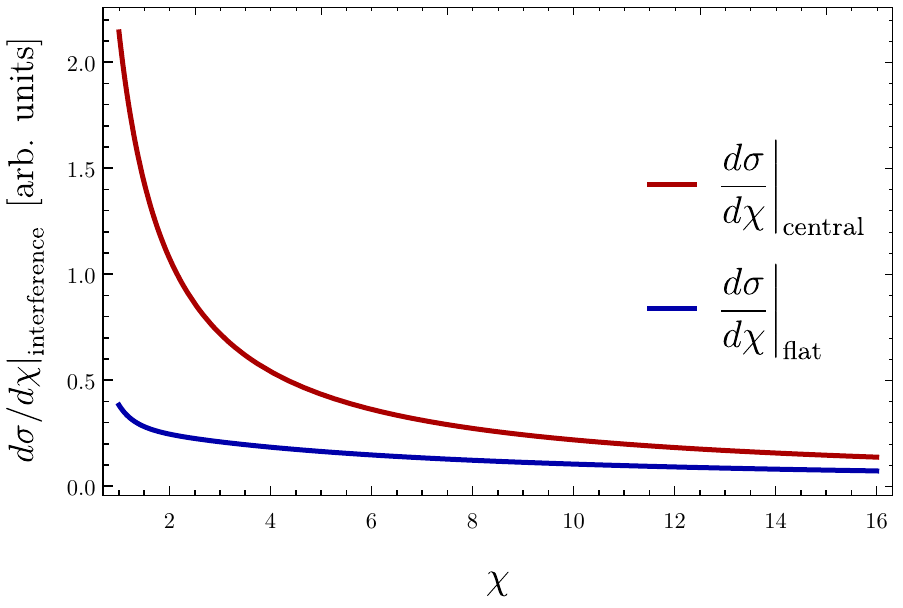}
 \caption{Parton-level distributions of the interference contributions due to the two distinct linear combinations of Wilson coefficients described in the text, shown in arbitrary units. These have been plotted for identical negative Wilson coefficients, giving a constructive interference effect.}
 \label{fig:partdist}
\end{figure}

The leading effect of BSM physics within the SMEFT is due to interference of the higher-dimensional operator amplitude with the QCD amplitude for dijet production. In principle, the effect of each operator could have distinct distributions in $\chi$, allowing for the measurement of such an angular spectrum to disentangle the contributions of multiple operators to dijet production. However, we find that within the SMEFT there are only two distinct shapes in the scattering angle which can be generated through interference. Thus, we may express the contribution from the interference to the differential cross section as
\begin{align}
\frac{1}{\Lambda^2} \left. \frac{d\sigma}{d\chi} \right|_\mathrm{interference} = \frac{1}{\Lambda^2} \left[ \left. \frac{d\sigma}{d\chi} \right|_\mathrm{central} + \left. \frac{d\sigma}{d\chi} \right|_\mathrm{flat}\right] \,.
\end{align}
For the parameters used later in our recast of the CMS analysis in Section~\ref{sec:single}, we show these two distributions at the parton level in Figure \ref{fig:partdist} by plotting the contributions proportional to the Wilson coefficients $c_{qq}^{(1)}$ and $c_{qu}^{(8)}$. The more centrally peaked SMEFT contribution to dijet production exhibits the following approximate dependence on the Wilson coefficients:
\begin{align}
\left. \frac{d\sigma}{d\chi} \right|_\mathrm{central}\propto
&- \left( 
c_{qq}^{(1)}
+0.61\,c_{qq}^{(3)}
+0.85\,c_{uu}
+0.15\,c_{dd}
+0.20\,c_{ud}^{(8)}
\right)  \,,\label{eq:sigmaCentral}
\end{align}
and the flatter distribution in $\chi$ is approximately proportional to
\begin{align}
\left. \frac{d\sigma}{d\chi} \right|_\mathrm{flat}\propto
- \left( 
c_{qu}^{(8)}
+ 0.45  c_{qd}^{(8)}
\right).\label{eq:sigmaFlat}
\end{align}
The coefficients displayed in front of the various Wilson coefficients vary with the center-of-mass energy of the process due to the differing behavior of up- and down-type quarks in the parton distribution functions. We quantify the ranges of these variations in Table~\ref{tab:wilsonVariations}; we anticipate that they will be too small to be exploited experimentally. To be concrete, we will utilize samples generated by turning on $Q_{qq}^{(1)}$ and/or $Q_{qu}^{(8)}$, the strongest contributors to these two angular distributions.

\begin{table}
\renewcommand{\arraystretch}{1.4}
\centering
\begin{tabular}{r|cccc|c}
\hline\hline
Wilson coefficient & $c_{qq}^{(3)}$ & $c_{uu}$ & $c_{dd}$ & $c_{ud}^{(8)}$  & $c_{qd}^{(8)}$ \\
\hline 
Range of prefactor & $[0.54, 0.67]$ & $[0.78, 0.91]$ & $[0.09, 0.22]$ & $[0.14, 0.26]$ & $[ 0.28, 0.64]$\\
\hline\hline
\end{tabular}
\caption{Range of the prefactor of the respective Wilson coefficients in equations~\eqref{eq:sigmaCentral} and~\eqref{eq:sigmaFlat}. Note that they are always normalized against the contribution from $c_{qq}^{(1)}$ in~\eqref{eq:sigmaCentral} and against $c_{qu}^{(8)}$ in~\eqref{eq:sigmaFlat}.}\label{tab:wilsonVariations}
\end{table}

It is possible to interpret the bounds coming from these searches in two different ways on a conceptual level. Of course, the signal size due to any operator is dependent only on the combination $c_k/\Lambda^2$. As a result, neglecting renormalization effects, any EFT search na\"\i vely only has access to this combination. This can be interpreted as either bounding the Wilson coefficient $c_k$ for an assumed NP scale $\Lambda$, or as a bound on the mass scale of NP $\Lambda$ assuming some coupling strength $c_k$. While these are equivalent statements for the signal strength predicted, they require distinct treatment when it comes to considering the theoretical errors inherent in the signal prediction.

\section{Consistency requirements for EFT analyses}
\label{sec:consistency}

As always in non-renormalizable EFTs, a new perturbation series that was not present in the SM has been introduced to the SMEFT. While there has been some discussion of the correct methodology for this power counting in general EFT extensions of the SM \cite{Buchalla:2013eza,Jenkins:2013sda,Buchalla:2014eca,Gavela:2016bzc,Buchalla:2016sop}, within the SMEFT and in absence of UV assumptions the appropriate expansion is clear; one simply expands in inverse powers of the NP scale $\Lambda$. This expansion is already explicit in the Lagrangian of \eqref{eq:lagrangian}, and in order for the EFT to have been treated in a fully consistent way it must be respected throughout the calculation.

Including contributions from dimension-six four-fermion operators, we can express the differential cross section ${d\sigma}/{d\chi}$ as an expansion in terms of ${1}/{\Lambda^2}$ according to
\begin{align}
 \frac{d\sigma}{d\chi} = \left.\frac{d\sigma}{d\chi}\right|_\mathrm{SM} + \frac{1}{\Lambda^2} \left.\frac{d\sigma}{d\chi}\right|_\mathrm{interference} + \frac{1}{\Lambda^4} \left.\frac{d\sigma}{d\chi}\right|_\mathrm{BSM} + \dots\,,
 \label{eq:lam}
\end{align}
where the first term arises from the square of the SM amplitude, the second term corresponds to the interference of the SM amplitude with the dimension-six amplitude and the third term denotes the contribution from the squared dimension-six amplitude. The dots indicate contributions from the unspecified EFT operators of dimension eight and higher. The challenge to consistency arises from the fact that those operators which are dimension eight contribute at the same order in $\Lambda$ as the BSM term above, which implies that keeping that term while neglecting the contributions arising from new operators at dimension eight is not a consistent use of perturbation theory. This treatment also yields results which depend on the choice of dimension-six basis, as the field redefinitions necessary to move from one basis to another generate shifts in the neglected dimension-eight parameters. While the utility of keeping the BSM term has been asserted in certain UV scenarios \cite{Biekoetter:2014jwa,Biekotter:2016ecg}, such a statement cannot be made in a model-independent way.

This is akin to performing an NLO computation and then choosing arbitrarily to keep the square of the virtual correction in addition to the consistent combination of the interference of the virtual correction with the leading-order amplitude and the appropriate real corrections. The reason that this inconsistency has persisted in the EFT literature while it would be considered unconscionable in e.g. higher order QCD calculations is that there is no IR divergence present in these amplitudes that manifestly must be canceled consistently order by order between real and virtual contributions. Thus, the squared contribution in this case, while not being consistent with the power counting of the perturbation theory, is at least not as manifestly ill-defined as it would be in the case of loop corrections with massless particles.

The two consistent, UV-independent ways of addressing this issue in EFT analyses are to either select a basis of dimension-eight operators and include their interference with the SM amplitude as part of the signal, thus consistently including all effects up to order $1/\Lambda^4$, or to drop the contribution due to the squared dimension-six amplitude as being a higher-order correction. In either case, an estimate of the size of the neglected terms in the perturbation series is needed to treat the data consistently, allowing interpretations of these bounds directly as constraints on NP models through matching to the EFT. Given the challenge of constraining the full set of operators at dimension six, expanding the analysis to include dimension-eight operator effects seems unlikely to improve our ability to constrain the parameter space; thus, we choose to follow the second path to a consistent EFT analysis here.

The state-of-the-art analyses from the ATLAS and CMS collaborations \cite{Aaboud:2017yvp,Sirunyan:2017ygf} do include all three specified contributions to the differential cross section in \eqref{eq:lam} as signal. As we just discussed, in principle one should either include the contribution of dimension-eight operators or truncate the series in \eqref{eq:lam} after the second term. In our study, we pursue the latter option and include terms up to order $1/\Lambda^2$ in the differential cross section. We take the BSM squared terms of order $1/\Lambda^4$ as a template for the distribution of the errors due to the neglect of the full class of new contributions arising at that order.  

\section{EFT searches in normalized angular distributions}
\label{sec:single} 

We shall now focus on a recent CMS analysis \cite{Sirunyan:2017ygf} for a center-of-mass energy of $\sqrt{s}=13\text{ TeV}$ and an integrated luminosity of $2.6 \text{ fb}^{-1}$. In this analysis, bounds on the NP scale $\Lambda$ were obtained for a fixed Wilson coefficient and considering the contribution of a flavor-diagonal color singlet operator. The analysis utilizes six bins in the invariant mass $m_{jj}$ of the pair of the two hardest jets in the process ranging from $m_{jj} = 1.9 \text{ TeV}$ up to $m_{jj}=13.0\text{ TeV}$. Bounds are derived by fitting angular differential distributions ${1}/{\sigma}~{d\sigma}/{d\chi}$, normalized to the total cross section in a given dijet mass bin. The bounds on $\Lambda$ are obtained by comparing data to theory predictions including NLO QCD and EW corrections, with factorization and renormalization scales chosen as the average transverse momentum $p_T$ of the two hardest jets. The fit is performed by applying a modified frequentist approach~\cite{Junk:1999kv,Read:2002hq}.

In our reproduction of the CMS analysis, we use Monte-Carlo samples for the three different pieces of the distribution ${1}/{\sigma}~{d\sigma}/{d\chi}$ from equation \eqref{eq:lam} generated at LO in QCD with \texttt{MadGraph5 v.~2.5.3} \cite{Alwall:2014hca}. We use \texttt{PYTHIA 6} \cite{Pythia1} for showering. For detector simulation, we employ \texttt{DELPHES v.~3.4.0} \cite{deFavereau:2013fsa}. We implemented the relevant SMEFT operators as a \texttt{FeynRules} model \cite{Alloul:2013bka}. We keep both the factorization and the renormalization scale fixed in every $m_{jj}$ bin. The fixed scales in the different $m_{jj}$ bins are determined by using Monte-Carlo samples of partonic events generated with \texttt{MadGraph5} at LO in QCD as follows: first, we generated a sample with a dynamical factorization and renormalization scale set to the average $p_T$ of the two partons in the final state. Afterwards, we determined the fixed scale for every $m_{jj}$ bin such that the cross section integrated over the whole range of $\chi$ agrees with the results from the runs with a dynamical scale and generate the production data sample utilizing that fixed scale for each bin; this better reproduces the angular dependence of the NLO results reported by CMS than a LO calculation with dynamical scales.

\subsection{Reproducing the CMS result}
\label{sec:CMS_repr}

\begin{figure}
\centering
\includegraphics[width=.49\textwidth]{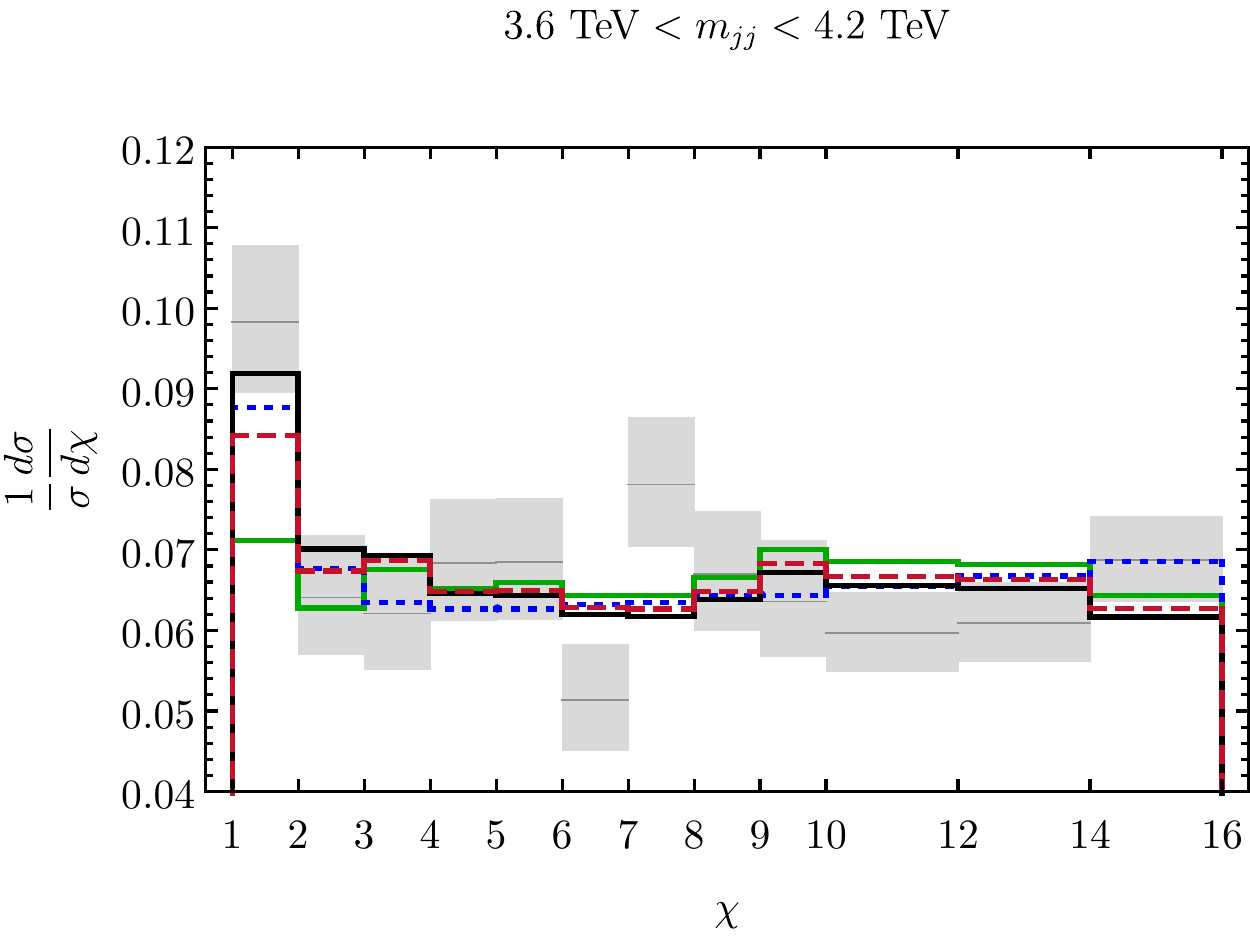}\hspace*{.01\textwidth}
\includegraphics[width=.49\textwidth]{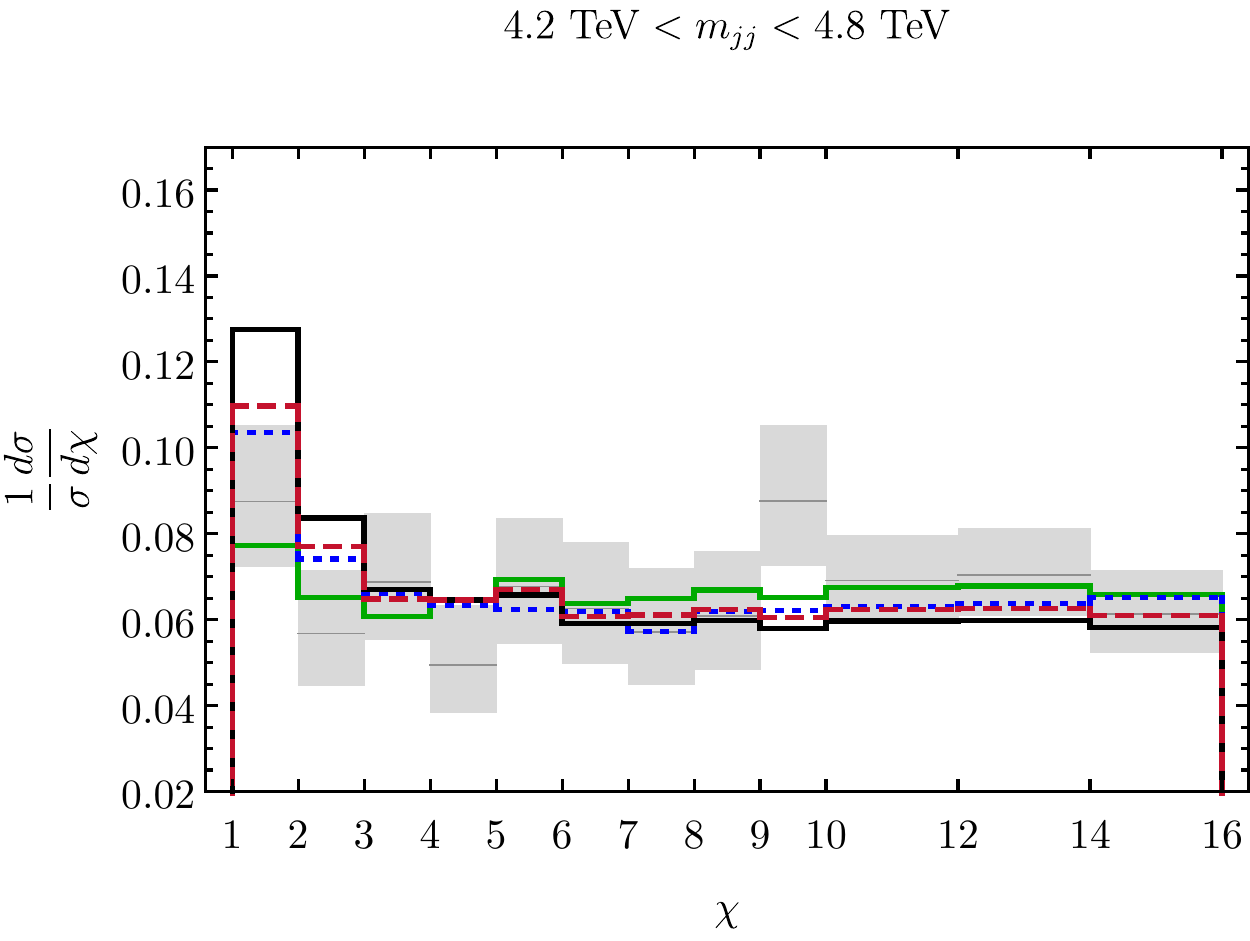}\\
\includegraphics[width=.49\textwidth]{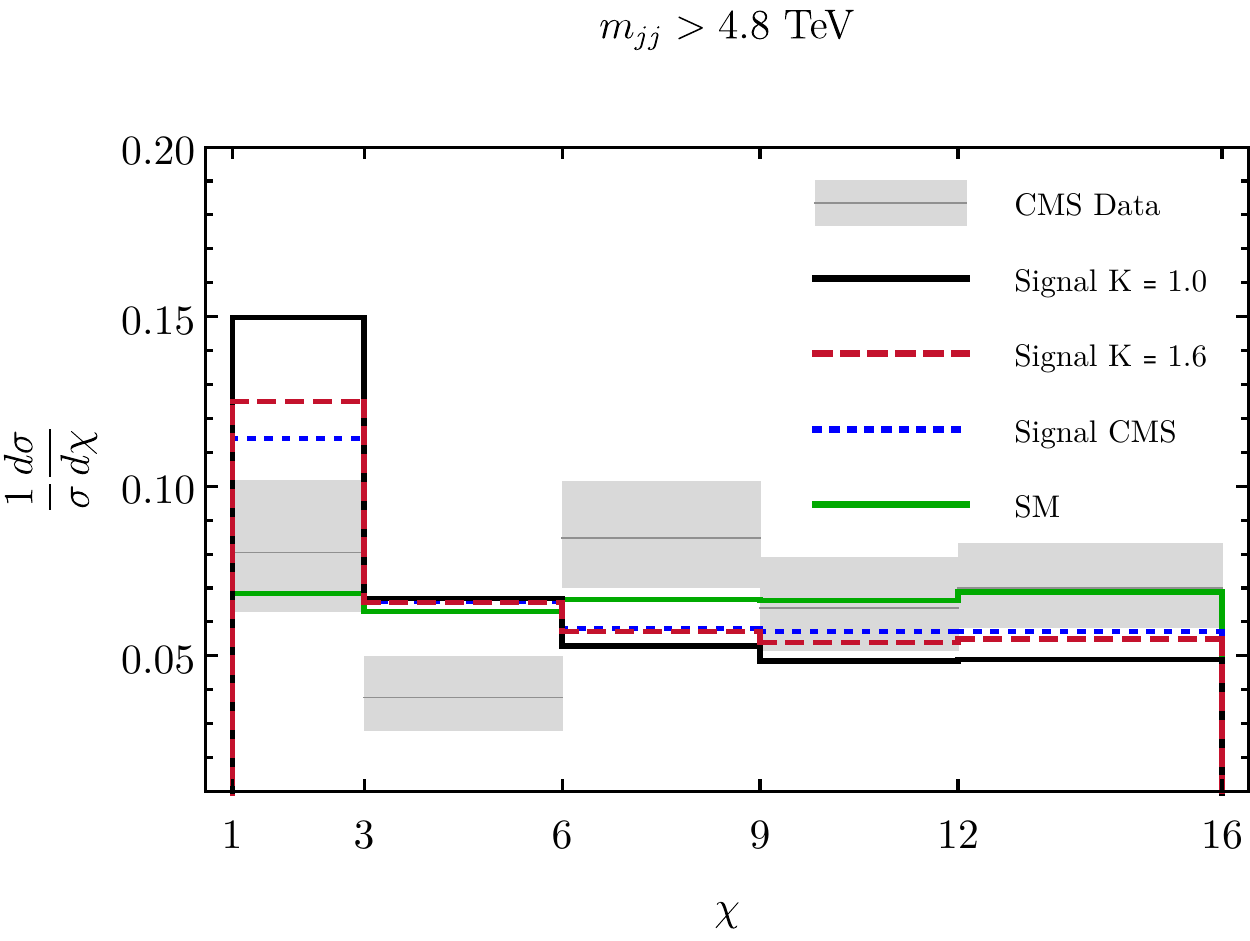}
\caption{Reproduction of the CMS analysis for $\Lambda=11 \text{ TeV}$ and $c_{qq}^{(1)}= + 2 \pi$. Shown are different normalized distributions in $\chi$ for the three highest bins in $m_{jj}$: the data with statistical and systematic uncertainties added in quadrature (gray rectangles), our signal for a background K-factor $K=1.0$ (black, solid), $K=1.6$ (red, dashed) and the CMS signal (blue, dotted). The green solid line depicts our SM signal prediction.}
\label{fig:CMS_recast}
\end{figure}

As a first step, and to validate our method, we perform the same analysis as the CMS collaboration \cite{Sirunyan:2017ygf} for the operator $Q_{qq}^{(1)}$, and derive bounds on $\Lambda$ for two different Wilson coefficients: the case $c_{qq}^{(1)}= + 2 \pi$ corresponds to destructive interference whereas $c_{qq}^{(1)}=-2\pi$ corresponds to constructive interference. We treat all three pieces in the distribution \eqref{eq:lam} as signal, in accordance with CMS's analysis. To compare our LO QCD results with the NLO QCD results from CMS, we allow for a rescaling of the SM piece of the signal by a factor $K$ in every $m_{jj}$ bin. We vary the $K$-factor between $1.0$ and $1.6$. The distributions in $\chi$  for the three highest $m_{jj}$ bins in the case $c_{qq}^{(1)} =+ 2 \pi$ and a scale $\Lambda=11\text{ TeV}$ are shown in Figure~\ref{fig:CMS_recast}; the remaining bins, lower in $m_{jj}$, do not provide an appreciable amount of constraining power, and so we neglect them.
We perform a Chi-squared test and fit the CMS data shown in gray in Figure~\ref{fig:CMS_recast}, adding the systematic and the statistical errors reported by CMS in quadrature. Our lower bounds for the NP scale are shown in Table~\ref{tab:repr_CMS}. All bounds derived in this work correspond to the 95\%~CL.
\begin{table}
\centering
\begin{tabular}{c|c|c}
 Wilson coefficient & $c_{qq}^{(1)}=+2\pi$ & $c_{qq}^{(1)}=-2\pi$ \\  
 \hline
 CMS  & 11.5 & 14.7 \\
 $K=1.0$ & 12.1 & 15.2 \\
 $K=1.3$ & 11.4 & 14.0 \\
 $K=1.6$ & 11.0 & 13.2 
\end{tabular}
\caption{Lower bounds for the NP scale $\Lambda$ in TeV obtained by considering the contribution of the SMEFT operator $Q_{qq}^{(1)}$ for the two different Wilson coefficients $c_{qq}^{(1)}=+ 2\pi$ (destructive interference) and $c_{qq}^{(1)}=- 2\pi$ (constructive interference). Shown are the bounds reported by CMS \cite{Sirunyan:2017ygf} and the bounds from our reproduction of that analysis for three choices of the $K$-factor.}
\label{tab:repr_CMS}
\end{table}
For comparison, we show the bounds reported by CMS in the second row. For $K=1.0$ in the three $m_{jj}$ bins, our bounds agree with the CMS results at the level of 5\%. Varying the K-factors separately in the $m_{jj}$ bins among the values $K\in\{1.0,1.3,1.6\}$, we find bounds which in the most extreme case differ from the CMS bound by 10\%. We suspect that these remaining small differences arise from not including NLO SM backgrounds in full, our different treatment of the factorization and the renormalization scales, and (primarily) from our simplified fit method.

\subsection{Consistent EFT treatment of the CMS data}
\label{sec:CMS_corr}

In the remainder of the analysis, we shall perform the EFT expansion in the mathematically consistent way and treat the contribution proportional to $\Lambda^{-4}$ as a higher-order term. Therefore, our predicted dijet angular distribution is 
\begin{align}
 \left.\frac{d\sigma}{d\chi}\right|_\text{signal} = \left.\frac{d\sigma}{d\chi}\right|_\text{SM} + \frac{1}{\Lambda^2} \left.\frac{d\sigma}{d\chi}\right|_\text{interference}+\mathcal{O}\left( \frac{1}{\Lambda^4}\right)  \,.\label{eq:consistentTruncation}
\end{align}
The piece now no longer included is instead used as an estimate of the theory uncertainty:
\begin{align}
\Delta\left(\frac{d\sigma}{d\chi}\right)_\mathrm{theo} = \frac{1}{\Lambda^4}\left.\frac{d\sigma}{d\chi}\right|_\text{BSM} \,.
\label{eq:BSMasError}
\end{align}
This is the most straightforward way of parameterizing the error, neglecting contributions from dimension-eight amplitudes interfering with the SM ones, which are of the same order as $\Lambda^{-4}\,d\sigma/d\chi|_\mathrm{BSM}$.

\begin{figure}
\centering
\includegraphics[width=.49\textwidth]{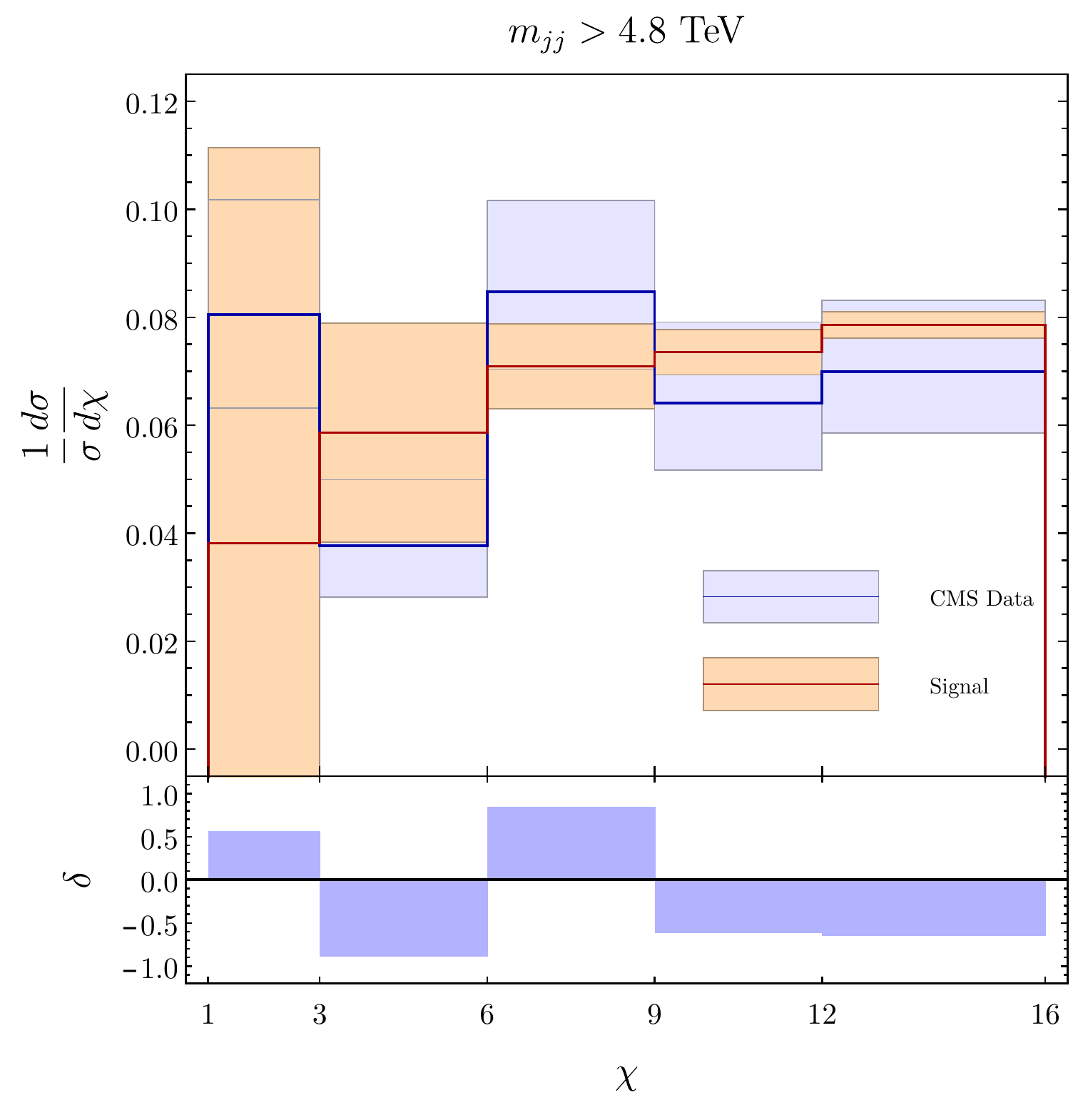}
\includegraphics[width=.49\textwidth]{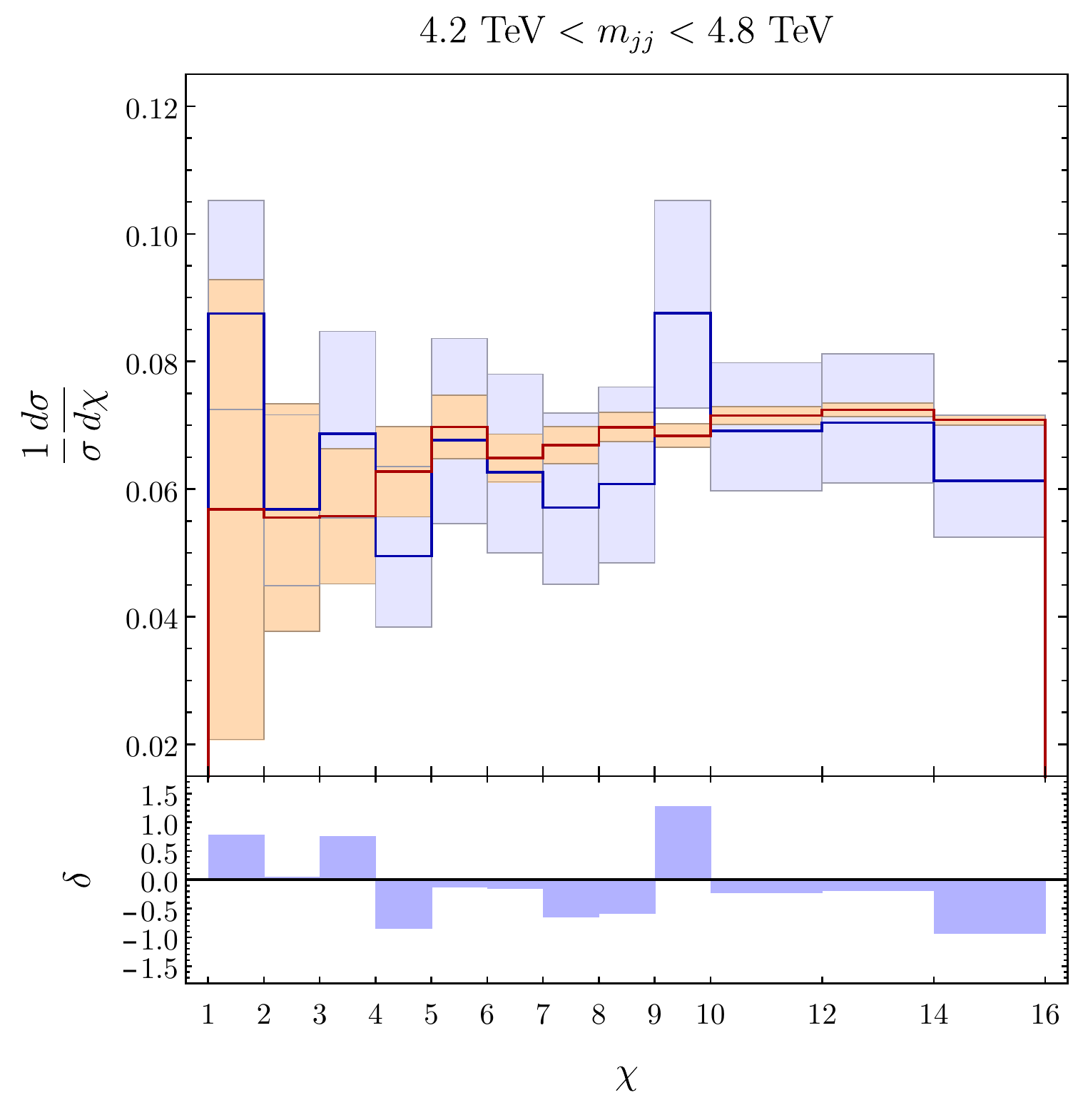}
\caption{Recast of the CMS analysis using consistent theoretical predictions truncating the EFT effects appropriately and including the newly introduced theory error. Here we consider $c_{qq}^{(1)}= + 2 \pi$ with $c_{qu}^{(8)}=0$ at a benchmark point $\Lambda=15~\mathrm{TeV}$. Shown is the data from CMS (blue, solid) with combined errors (blue-shaded region), the prediction when truncated consistently (red, solid) and the newly introduced theory errors (orange-shaded region). The bottom panels show the fit pulls $\delta$, defined as the difference between the central values divided by the combined errors.
}
\label{fig:CMS_recast_theorerr}
\end{figure}  
The distributions of the data in the two highest bins in the dijet invariant mass $m_{jj}$ are shown in Figure~\ref{fig:CMS_recast_theorerr}. In the upper panels, we show the CMS data as solid blue lines with the associated statistical and systematic errors added in quadrature as shaded-blue regions. The solid red lines  and the orange-shaded regions show the predictions~\eqref{eq:consistentTruncation} with $K=1.0$ as well as the associated theory error~\eqref{eq:BSMasError}, both normalized to the total cross section in each $m_{jj}$ bin. The bottom panels show the respective fit pulls, defined by
\begin{align}
 \delta = \frac{\mu_\mathrm{CMS} - \mu_\mathrm{signal}}{\sqrt{\Delta \mu^2_\mathrm{CMS}+\Delta \mu^2_\mathrm{theo}}}\,,
\end{align}
where $\mu$ is a shorthand notation for the differential cross section in each $\chi$ bin normalized to the total cross section in the corresponding $m_{jj}$ bin. A Chi-squared test\footnote{We note that a likelihood-ratio analysis would be more complete and statistically motivated given the theory errors we consider; lacking the experimental details necessary to build a realistic likelihood function we continue with Chi-squared analyses here.} utilizing these distributions yields no bound at all on the EFT, as is evident given the small values of the fit pulls in Figure \ref{fig:CMS_recast_theorerr}. This can be clearly attributed to the new treatment of the squared EFT piece as a theory error: while the signal was strongest and in the fit most constrained in the low $\chi$ bins, the signal is now more flat across the angular variable. Instead, the previously huge contribution to the signal is now part of the uncertainties and completely overwhelms the signal in those bins. Clearly, the search strategy needs to be altered to obtain meaningful bounds on the EFT parameters.

\section{EFT searches in unnormalized angular distributions}
\label{sec:full}

In this section, we perform the search in the angular distributions again, but without normalizing the differential cross sections in the different $\chi$ bins to the total cross section in the corresponding $m_{jj}$ bin. While the normalization is beneficial for cancelling uncertainties in the signal prediction and experimental reconstruction, it also removes part of the information contained in the data, which can potentially yield more constraints on the fit to place bounds on the EFT parameters even when consistently truncating the expressions.

We use a set of pseudodata generated in the same way as explained in Section~\ref{sec:single}. We now consider the presence of the two distinct linear combinations of operators by also turning on $Q_{qu}^{(8)}$. We treat the systematic error reported for the normalized distributions by CMS as a percentage and apply it to our event count, and introduce a factor $C_\text{syst}$ rescaling the systematic error reported by CMS to account for the likely underestimate of systematics in moving from normalized to unnormalized distributions. Note that we already assume that the systematics are not improved (in fractional terms) by the increase of data, so that our treatment for larger data sets is somewhat conservative from the start. For the statistical error we utilize a Poisson error for the number of events in every $\chi$-bin.

\begin{figure}
\centering
\includegraphics[width=0.495\textwidth]{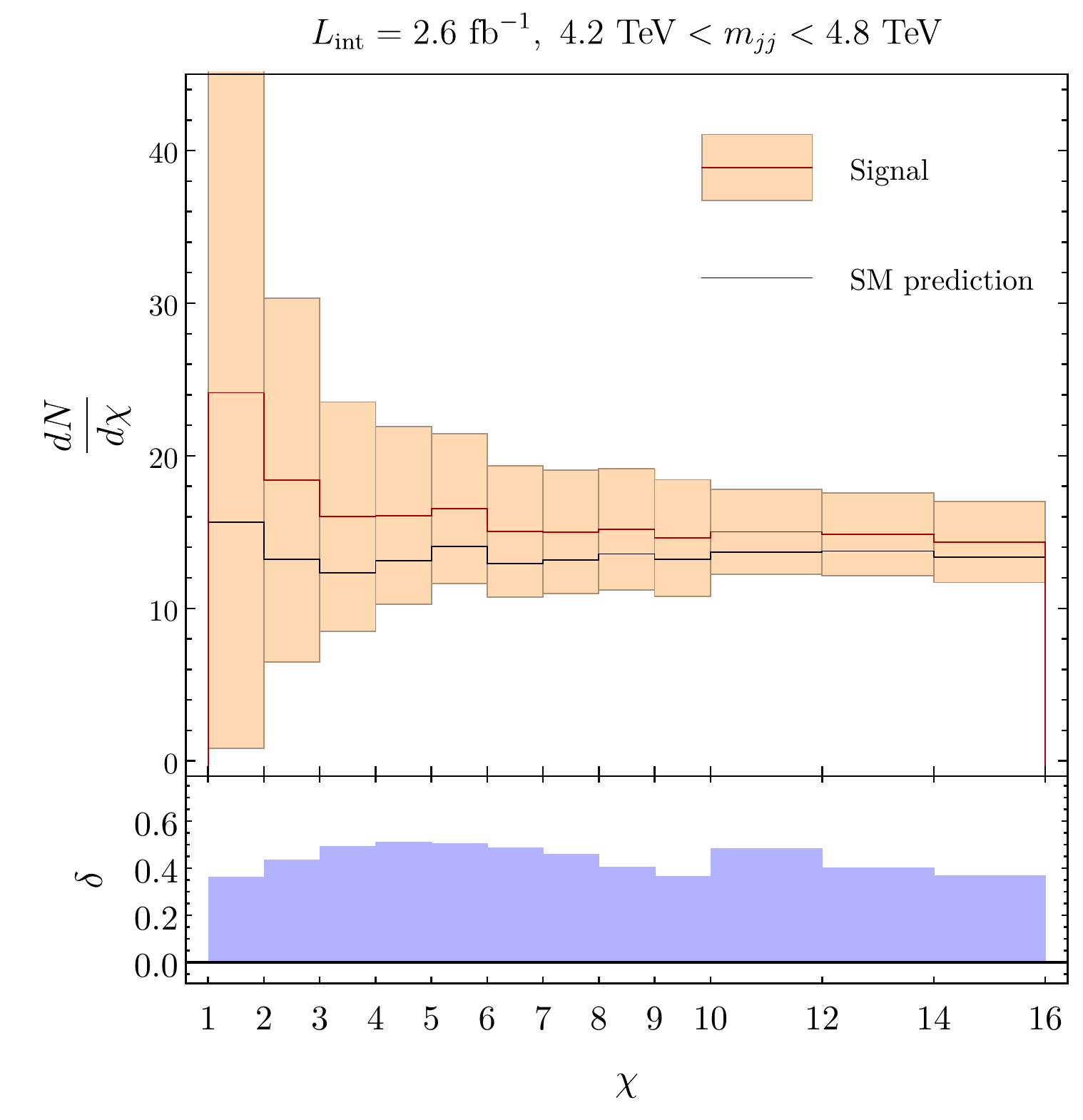}
\includegraphics[width=0.495\textwidth]{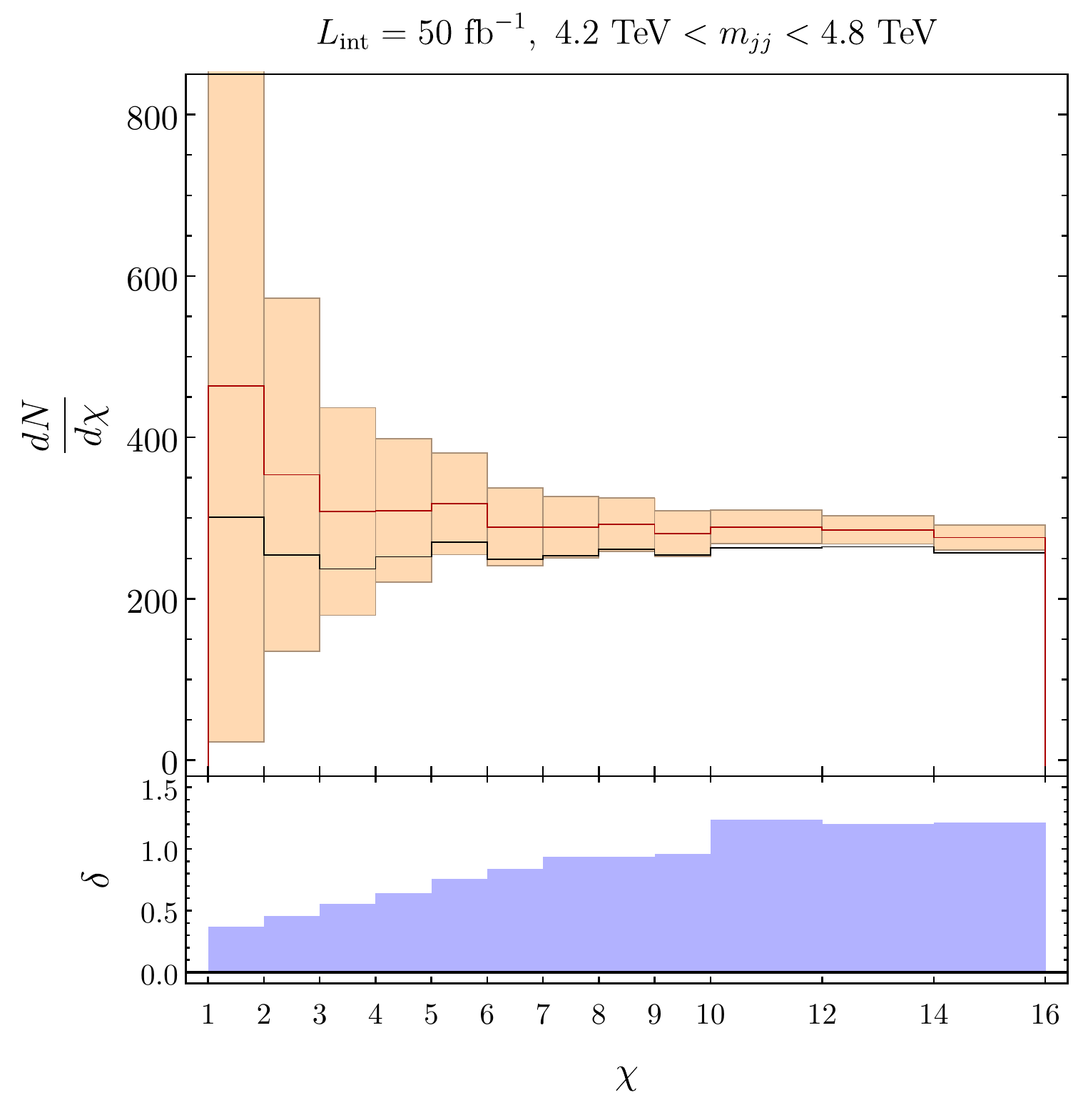}
\caption{Unnormalized angular distributions for $C_\text{syst}=1$ and the integrated luminosities \mbox{$L_\text{int}=2.6\text{ fb}^{-1}$} (left) and $L_\text{int}=50.0\text{ fb}^{-1}$ (right). Shown is the event count from SM pseudodata (black line) and the predicted signal for $\Lambda=11\text{ TeV}$, $c_{qq}^{(1)} = - 2 \pi$, and $c_{qu}^{(8)}=0$ with the new theory error, the statistical error and the rescaled systematic errors added in quadrature (red line and shaded region). The bottom panels show the corresponding fit pulls.
}
\label{fig:unnormchi}
\end{figure}

Our results for the unnormalized angular distributions are shown in Figure~\ref{fig:unnormchi} for \mbox{$C_\text{syst}=1$}, $c_{qq}^{(1)}=-2\pi$,  $c_{qu}^{(8)}=0$, the NP scale $\Lambda=11$~TeV, and the integrated luminosities \mbox{$L_\text{int}=2.6\text{ fb}^{-1}$} (left panel) and \mbox{$L_\text{int}=50.0\text{ fb}^{-1}$} (right panel). 
It is important to note that the current analysis is mainly limited by statistics for $C_\text{sys}=1$. Whereas the relative error is sizeable in the lowest bins in $\chi$ also for $\mathcal{L}_\text{int}=50 \text{ fb}^{-1}$, it is strongly reduced for larger $\chi$ by increasing the luminosity. We exploit these unnormalized distributions to derive bounds on NP scenarios in two different ways using a Chi-Squared fit, as discussed in Section~\ref{sec:SMEFT}. We either fix the Wilson coefficients and obtain bounds on the cutoff scale of the EFT, or fix the scale and constrain the Wilson coefficients.

We model the theory error initially as the sum of the two contributions arising from squared EFT amplitudes of the operators $Q_{qq}^{(1)}$ and $Q_{qu}^{(8)}$; these two operators cannot interfere due to their flavor and color structure. Due to the different combinatorics of their flavor and color structures, the error proportional to ${c_{qq}^{(1)}}^2$ is more than an order of magnitude greater than that proportional to ${c_{qu}^{(8)}}^2$. We then introduce two models for the theory error to include unknown dimension-eight effects, based on the distribution of the squared BSM pieces in the angular variable $\chi$. In these distributions we replace each squared Wilson coefficient $c_k^2$ by a quantity $\Delta_{\mathrm{theo},i}$, which we define such that it can address the contribution from a dimension-eight amplitude interfering with the SM one. Our first model for said theory error is 
\begin{align}
\Delta_{\mathrm{theo},1} := \mbox{max}\left\{c_k^2  ; \, g_s \, c_8 \sqrt{N_8} \right\} \,,\label{eq:theoryError1}
\end{align}
where $g_s$ is the strong coupling, $c_8$ represents the Wilson coefficient of the dimension-eight operators and $N_8$ is the number of operators contributing at dimension eight. In this model we explicitly assume that one of the two sources of error contributing at order $1/\Lambda^4$ is dominant, and neglect the other. As an alternative model, we add both error contributions in quadrature, yielding
\begin{align}
\Delta_{\mathrm{theo},2} := \sqrt{c_k^4  + \left( g_s \, c_8 \sqrt{N_8} \right)^2 } \,.\label{eq:theoryError2}
\end{align}
For this search, we take $c_8=1/2\,\sum_k |c_k|$ so that the hypothetical dimension-eight Wilson coefficients are not arbitrarily assumed to be either much larger or smaller than those at dimension-six.

In the first part of this analysis, we use $C_\text{syst}=1$, consider different choices for the size of the Wilson coefficients $c_{qq}^{(1)}$ and $c_{qu}^{(8)}$, namely $0, \pm 1,\pm2 \pi$ and $\pm16\pi^2$, and derive bounds on the NP scale $\Lambda$. Since our analysis is mainly limited by statistics, there exists a minimum integrated luminosity which is necessary to obtain a bound. Above these minimal values of $L_\mathrm{int}$, we no longer simply find a lower bound on $\Lambda$ but rather a region of excluded values for $\Lambda$. This effect is caused by the theoretical error growing with the inverse fourth power of the scale, preventing us from deriving constraints at scales small compared to the dijet invariant masses giving the dominant contribution. Note that this is exactly what we should expect, and arises as well, though in an ad-hoc way, in studies which discard events at energies beyond the cutoff scale \cite{Englert:2014cva,Contino:2016jqw,Farina:2016rws,Alioli:2017jdo}.

\begin{figure}
 \centering
\includegraphics[width=.49\textwidth]{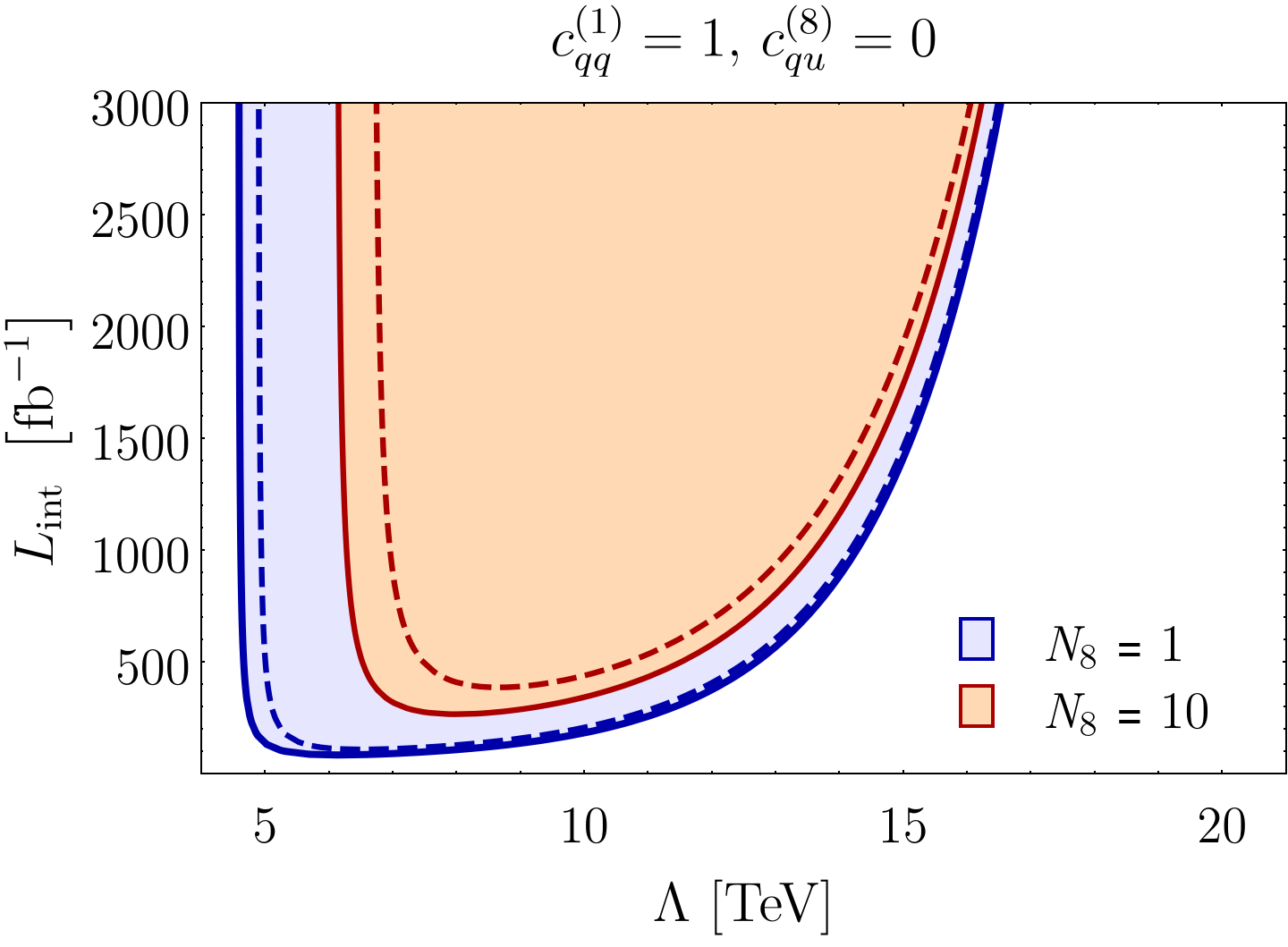}
\includegraphics[width=.49\textwidth]{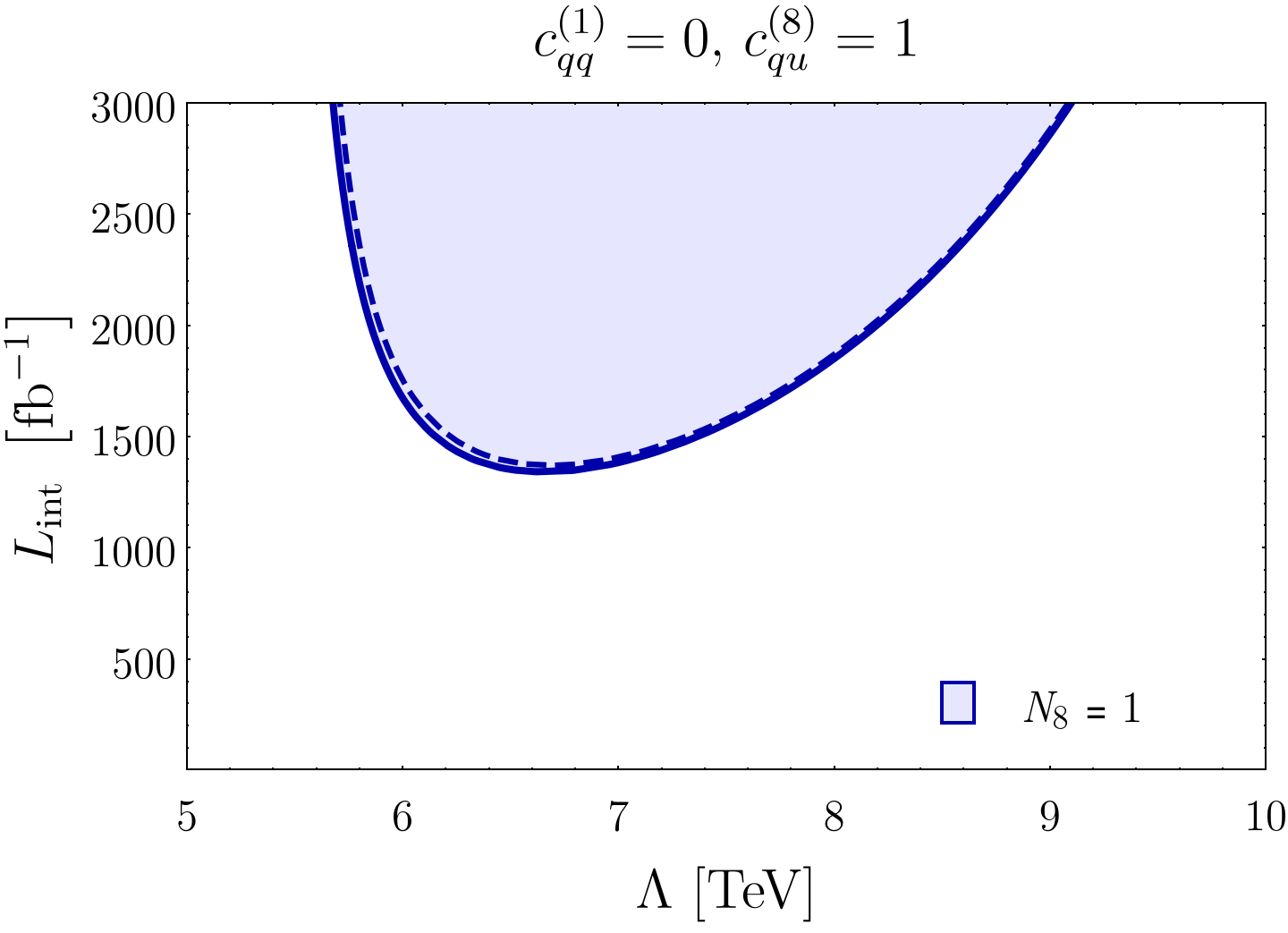}
\includegraphics[width=.49\textwidth]{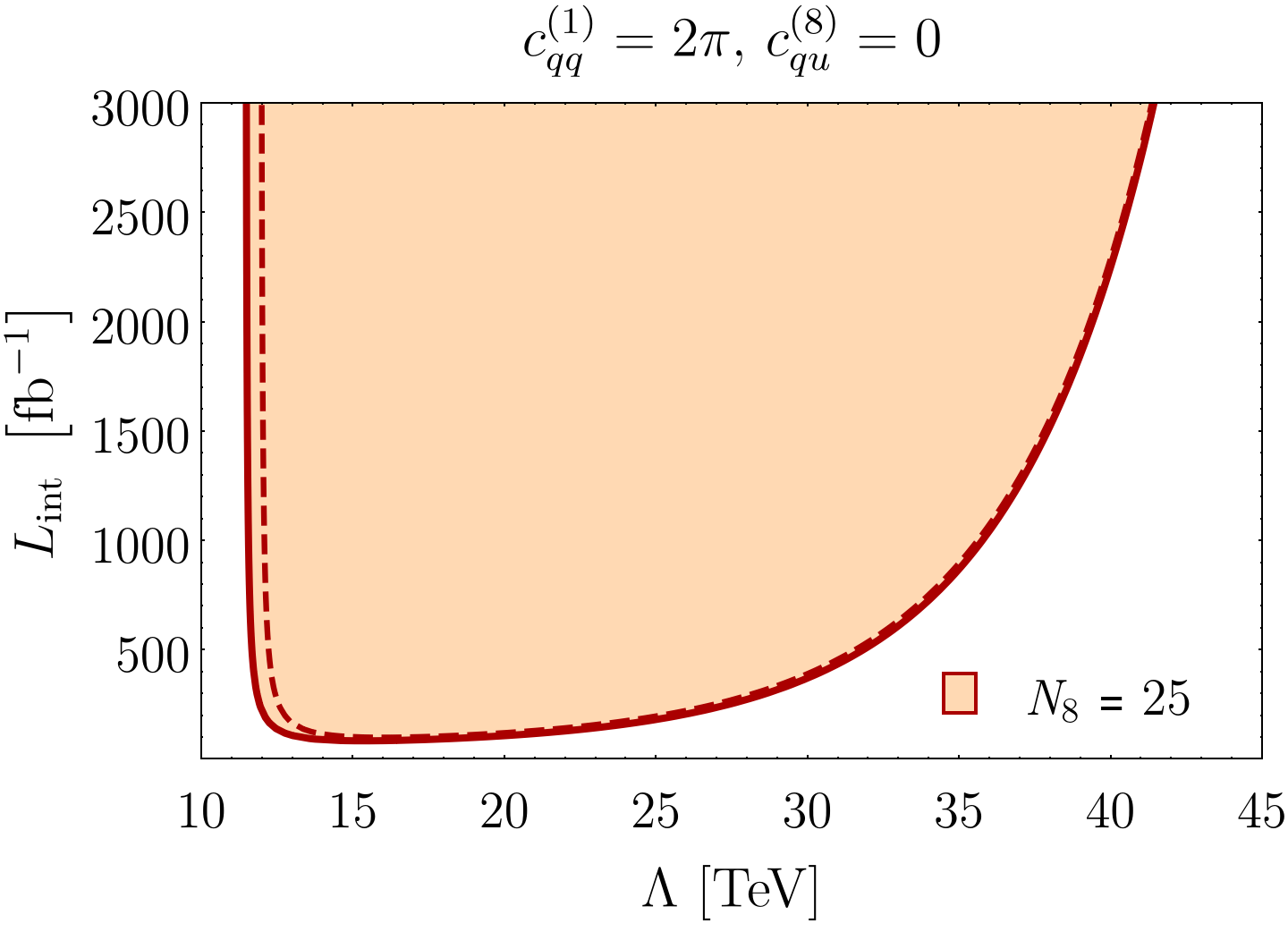}
\includegraphics[width=.49\textwidth]{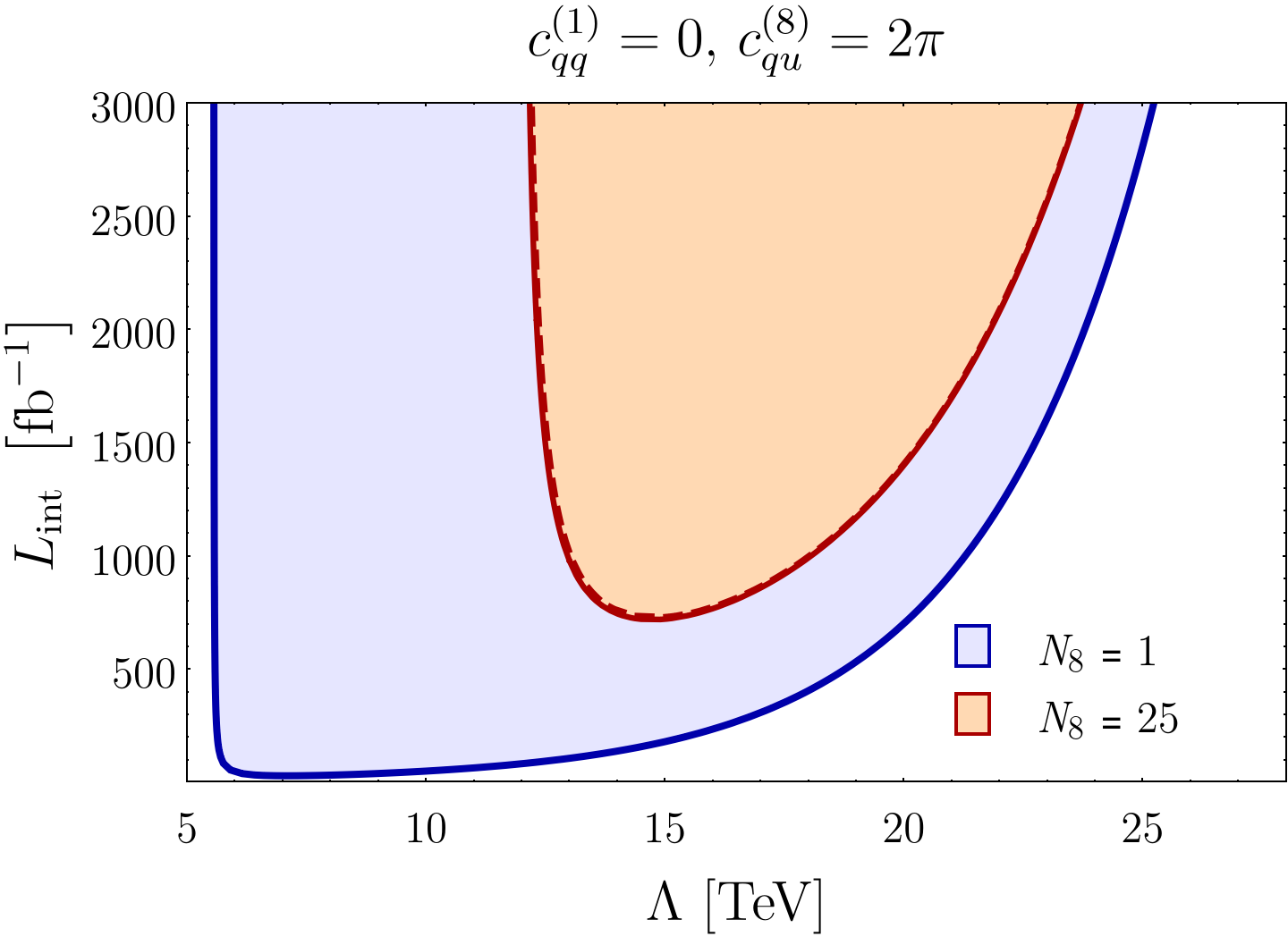}
\caption{Projected exclusion regions for the fixed-Wilson-coefficient case at different integrated luminosities for different choices for $c_{qq}^{(1)}$ and $c_{qu}^{(8)}$, where only a single coefficient is switched on at a time. The values chosen are $0, 1$ and $2 \pi$.
 The blue regions correspond to $N_8=1$ whereas the orange regions show the exclusions for $N_8=10$ or $N_8=25$ depending on the Wilson coefficients. Regions bounded by solid lines show the exclusions when the theory error $\Delta_{\mathrm{theo},1}$ is used, whereas dashed lines illustrate the exclusions when the more conservative error $\Delta_{\mathrm{theo},2}$ is used.}
 \label{fig:FW_theo1_a}
\end{figure}
\begin{figure}
 \centering
\includegraphics[width=.49\textwidth]{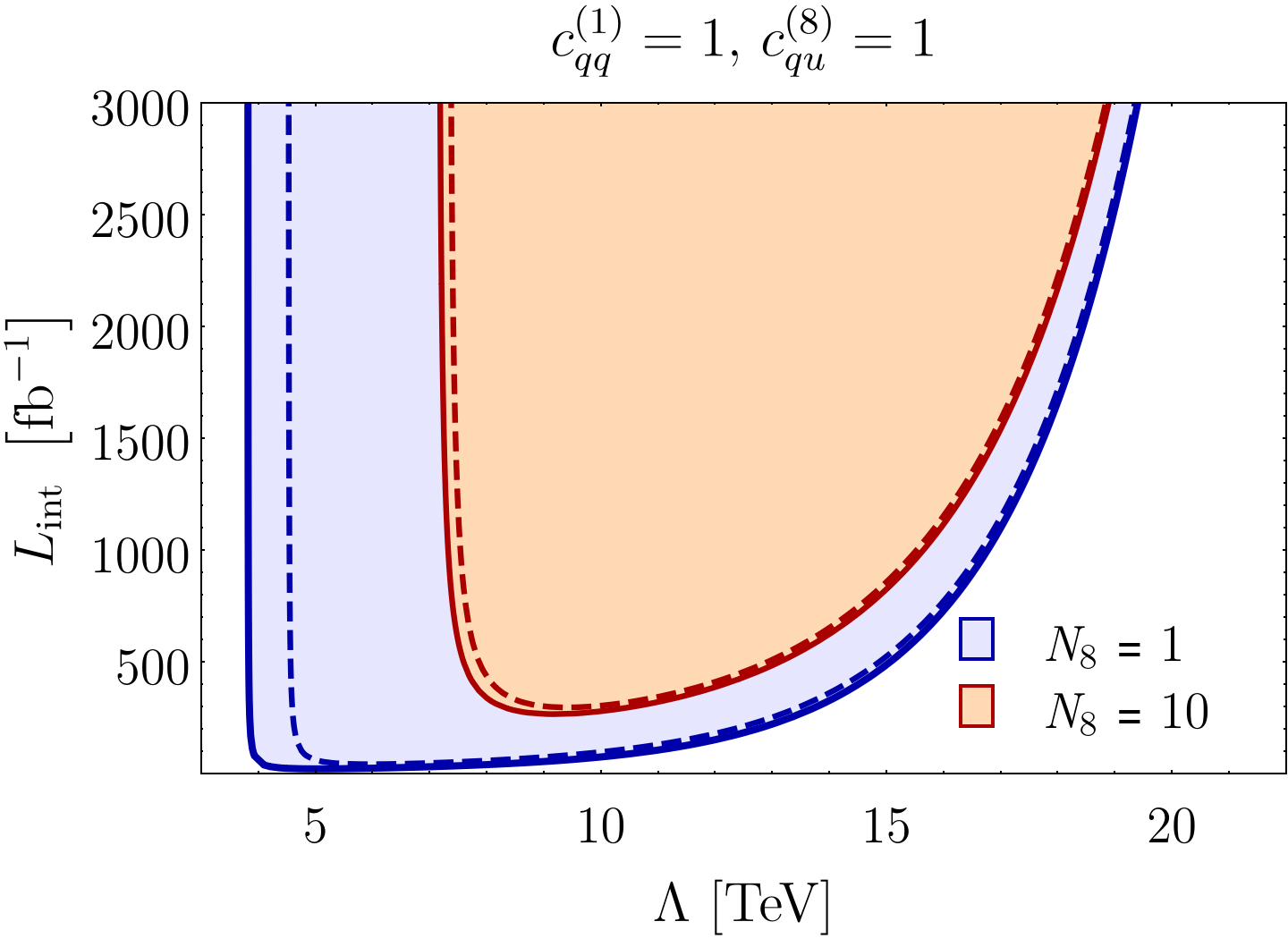}
\includegraphics[width=.49\textwidth]{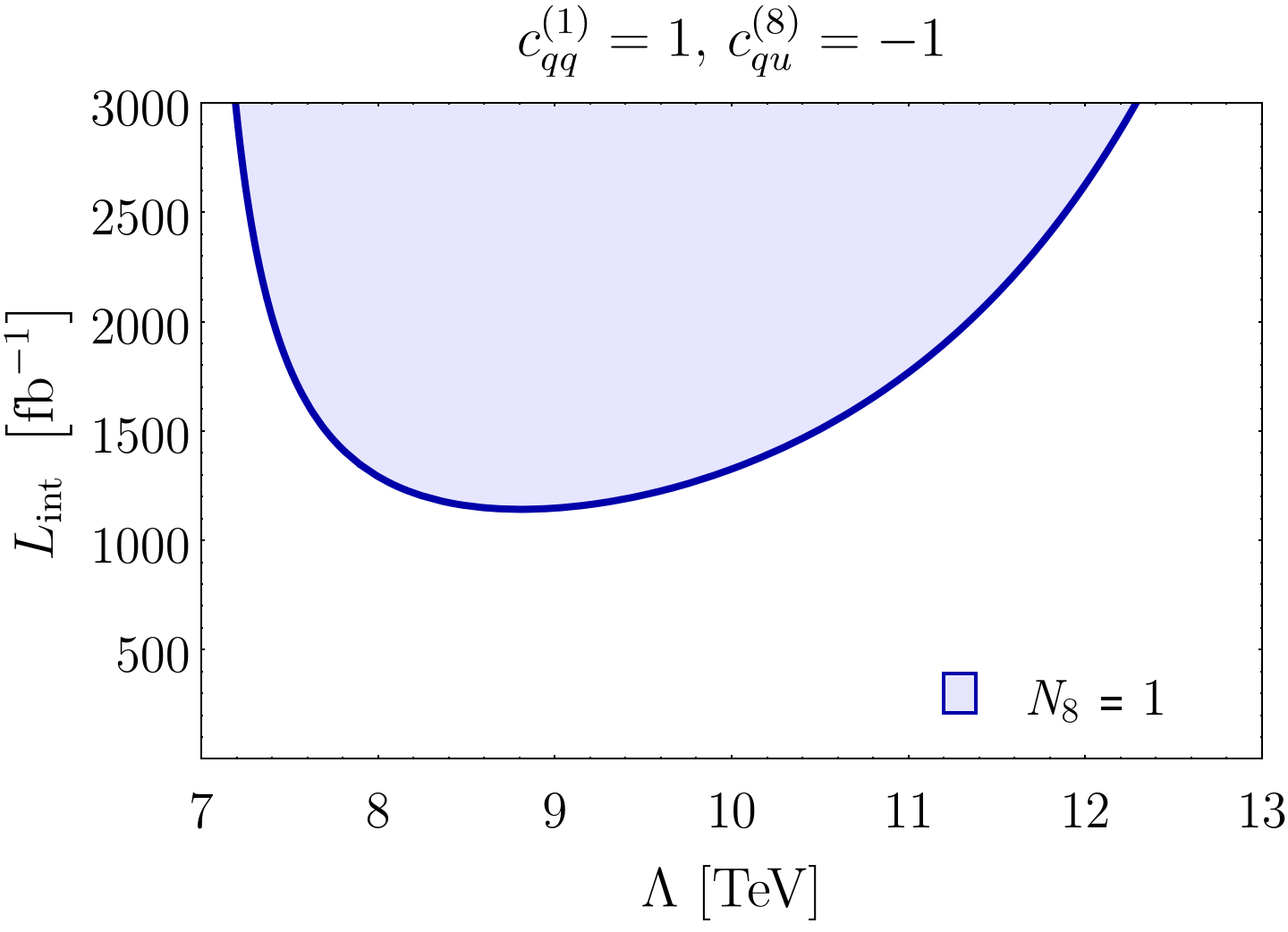}
\includegraphics[width=.49\textwidth]{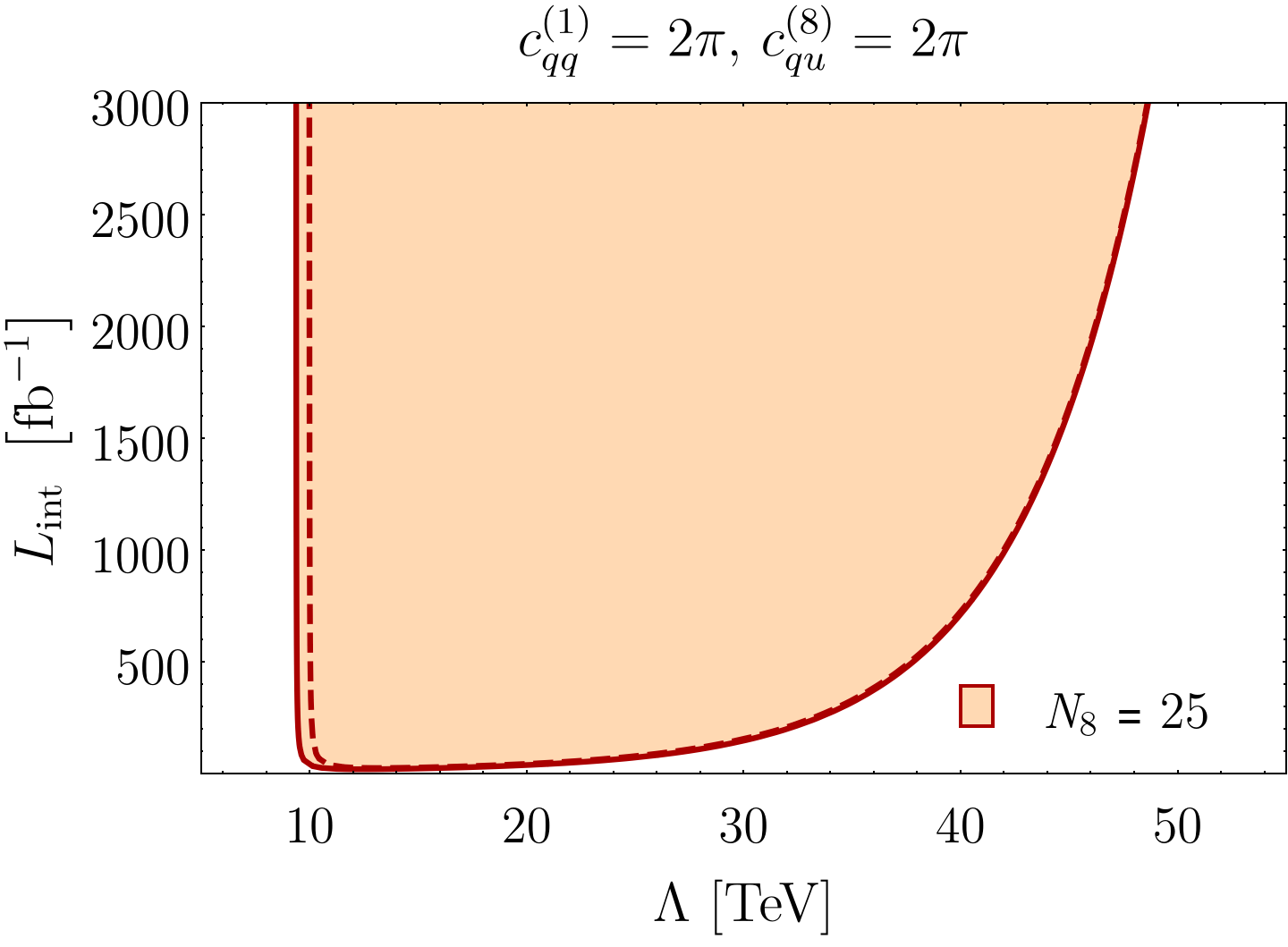}
\includegraphics[width=.49\textwidth]{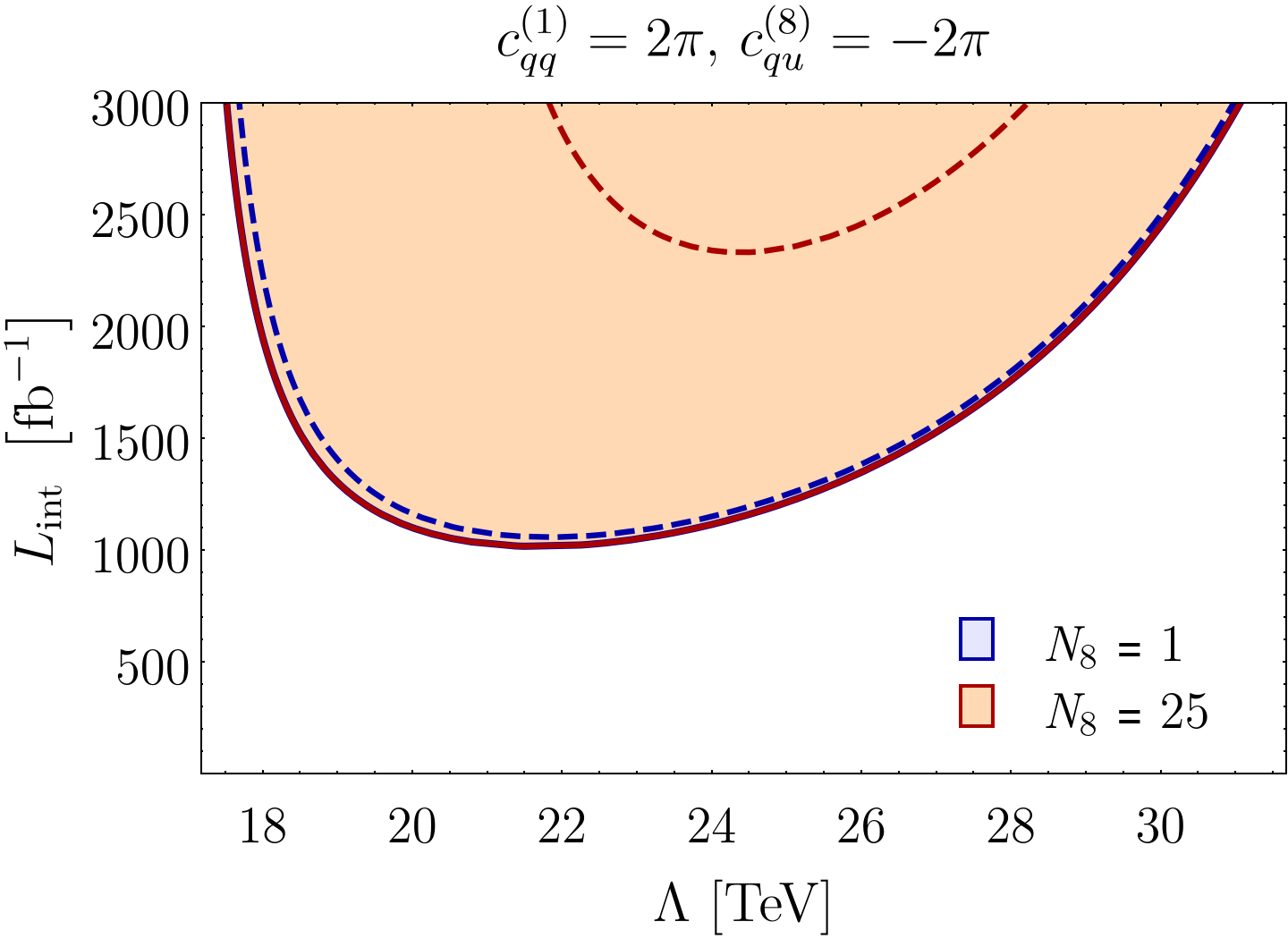}
\caption{Projected exclusion regions for the fixed-Wilson-coefficient case at different integrated luminosities for different choices for $c_{qq}^{(1)}$ and $c_{qu}^{(8)}$, where both coefficients are switched on simultaneously. The values chosen are $\pm1$ and $\pm2 \pi$.
 The blue regions correspond to $N_8=1$ whereas the orange regions show the exclusions for $N_8=10$ or $N_8=25$ depending on the Wilson coefficients. Regions bounded by solid lines show the exclusions when the theory error $\Delta_{\mathrm{theo},1}$ is used, whereas dashed lines illustrate the exclusions when the more conservative error $\Delta_{\mathrm{theo},2}$ is used.}
 \label{fig:FW_theo1_b}
\end{figure}

In Figures~\ref{fig:FW_theo1_a} and~\ref{fig:FW_theo1_b}, we show exclusion regions for the new-physics scale varying with integrated luminosity for different choices of the Wilson coefficients, namely $0, \pm 1,\pm2 \pi$. Figure~\ref{fig:FW_theo1_a} shows the case where only one coefficient is assumed to be non-zero at a time. Generically, one finds that for equal coupling strengths the bounds on the scale coming from $c_{qu}^{(8)}$ are weaker compared to the ones from $c_{qq}^{(1)}$, since the leading contribution to dijet production comes from $c_{qq}^{(1)}$. For $c_{qu}^{(8)}=1$ we observe that bounds can only be placed on the NP scale for signficant amounts of collected data. Even then, the excluded region is only a small window and when $N_8=10$ is chosen, no bounds can be derived at all. For the more strongly-coupled case where the coefficients are assumed to be $2\pi$, the exclusion regions grow and the minimum amount of integrated luminosity decreases. For $c_{qu}^{(8)}=2\pi$ it is possible to place bounds on $\Lambda$ even with larger values of $N_8$, although the exclusion regions shrinks drastically.

In Figure~\ref{fig:FW_theo1_b}, bounds are shown for the case where both coefficients contribute simultaneously. When both coefficients contribute with the same sign, the exclusion regions naturally grow compared to the single-Wilson case discussed above: now, larger ranges of $\Lambda$ can be excluded at lower integrated luminosities. When the two coefficients have opposite signs, their contributions partially cancel, leading to much weaker bounds that need more collected data to be derived.

Since we assume that the data matches the SM prediction exactly, the signs of the Wilson coefficients only play a role relative to each other. An overall sign change enters only through the slight differences in systematic errors on the high- and low-sides of the data, and change the limits only by small amounts. For all the cases presented in Figures~\ref{fig:FW_theo1_a} and ~\ref{fig:FW_theo1_b}, we observed a difference of less than 0.3 TeV in the constrained scale between overall sign choices. The impact of the choice for $N_8$ is most noticeable when the Wilson coefficients are of $\mathcal O (1)$. For large values of the Wilson coefficients, the theory errors $\Delta_{\mathrm{theo},i}$ are generally dominated by the piece independent of $N_8$.

For the maximally-coupled case, where the coefficients are set to $16\pi^2$, the analysis can, at an integrated luminosity of $1000\text{ fb}^{-1}$, only exclude the range $60~\mathrm{TeV}\lesssim\Lambda\lesssim 180~\mathrm{TeV}$ when only $c_{qq}^{(1)}$ is switched on, $15~\mathrm{TeV}\lesssim\Lambda\lesssim 110~\mathrm{TeV}$ when only $c_{qu}^{(8)}=16\pi^2$, and $50~\mathrm{TeV}\lesssim\Lambda\lesssim 210~\mathrm{TeV}$ when both coefficients are set to $16\pi^2$. The smallness of the lower bound for $\Lambda$ in the case $c_{qq}^{(1)}=0$ and $c_{qu}^{(8)}=16 \pi^2$ is caused by the fact that the prefactor of the dominant error distribution from $Q_{qq}^{(1)}$ contains the large Wilson coefficient only linearly.

We study the effect of varying the rescaling factor $C_\text{syst}$ by considering the case $c_{qu}^{(8)}=0$ and using the error $\Delta_{\text{theo},2}$. For this analysis we fix $N_8=10$ and determine the maximum of the factor $C_\text{syst}$ under the condition that one is still able to derive a bound with an integrated luminosity of 3000 $\text{fb}^{-1}$. We find the values $C_\text{syst}\approx 3,7,7$ corresponding to the Wilson coefficients $c_{qq}^{(1)}=1,2\pi,16\pi^2$. Especially in the latter two cases, a study of unnormalized angular distributions is thus reasonably robust with respect to systematic errors.

When the NP scale $\Lambda$ is fixed, bounds on the Wilson coefficients can be derived. The introduction of the new theory error has the same effect we observed already in the case of fixed Wilson coefficients: instead of an upper bound on the couplings, there is also a value of the Wilson coefficient above which the theoretical error grows too large to rule out the parameter point. Thus, we generate bounded ellipses in the $(c_{qq}^{(1)},\,c_{qu}^{(8)})$-plane rather than an allowed ellipse, as is more common. 

\begin{figure}
 \centering
 \includegraphics[width=.48\textwidth]{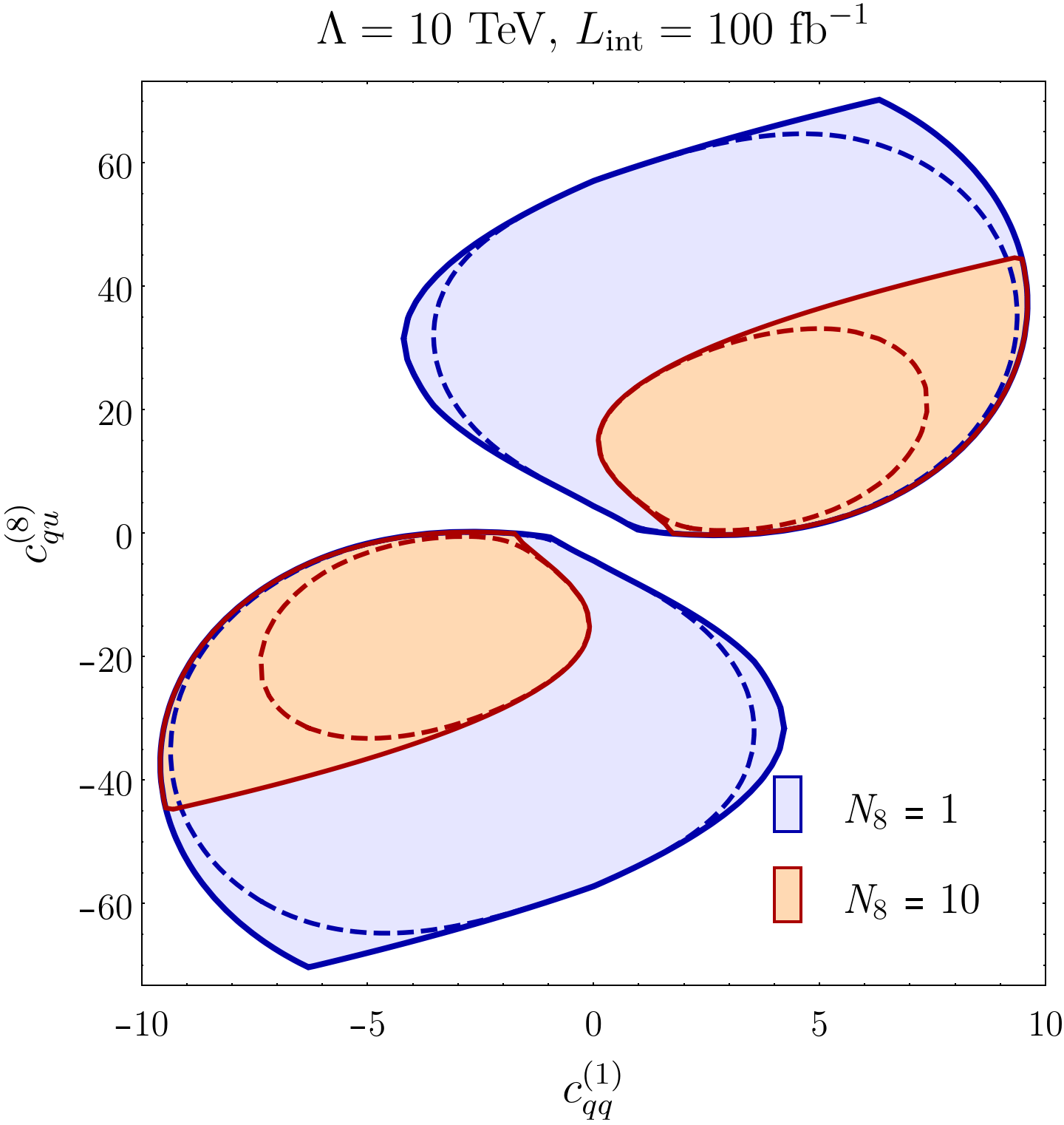}
 \includegraphics[width=.48\textwidth]{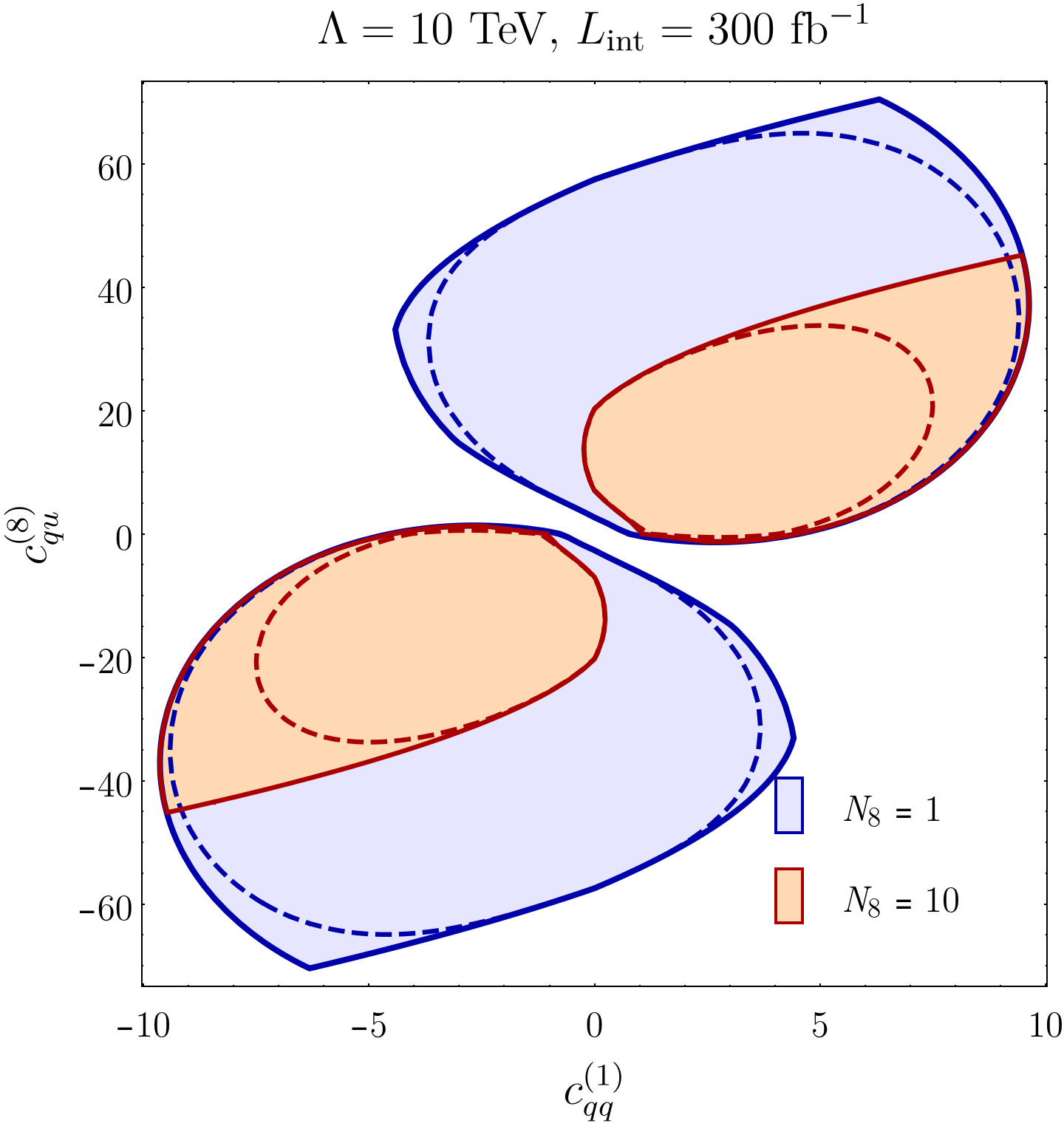}
 \includegraphics[width=.48\textwidth]{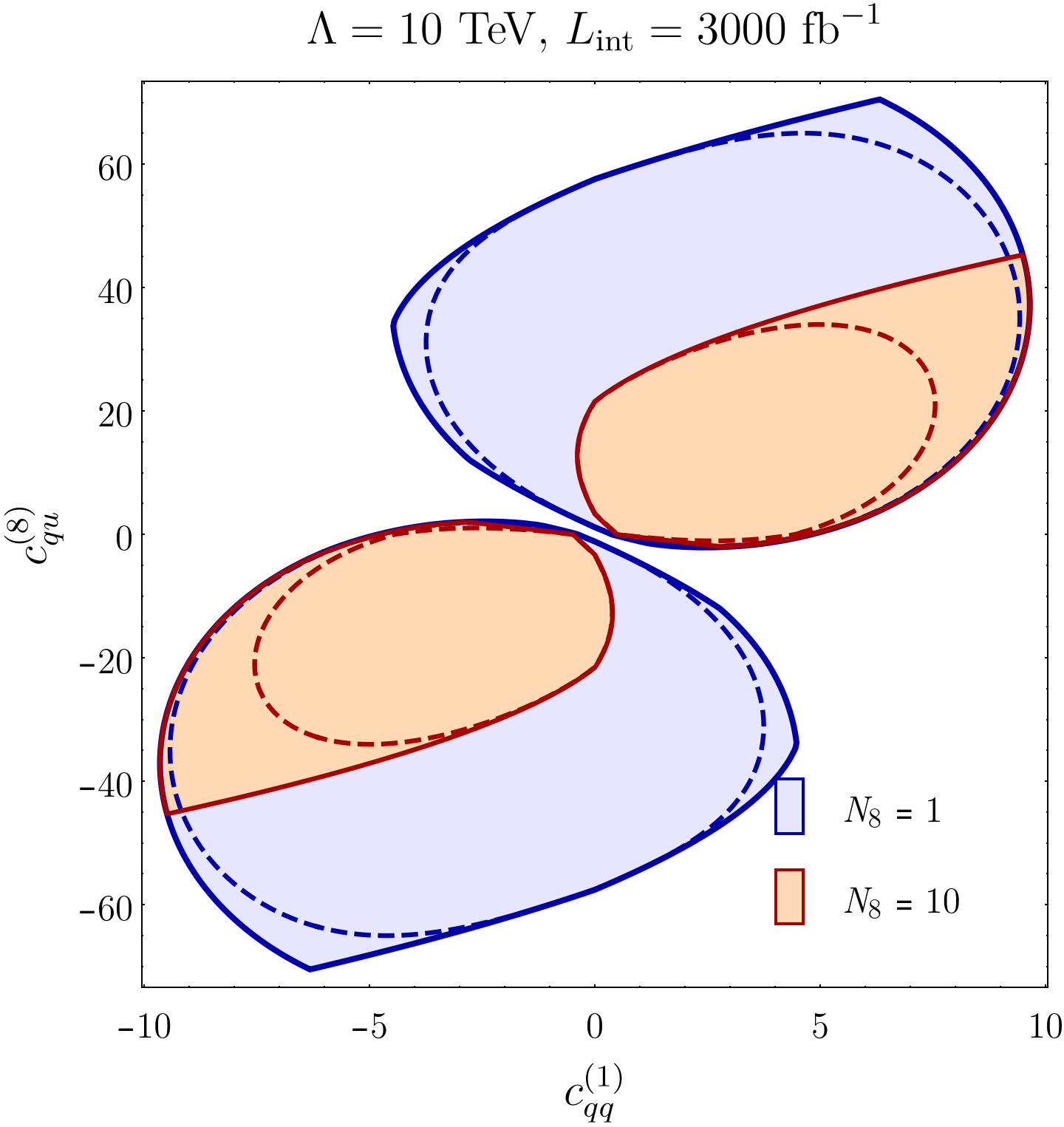}
 \includegraphics[width=.48\textwidth]{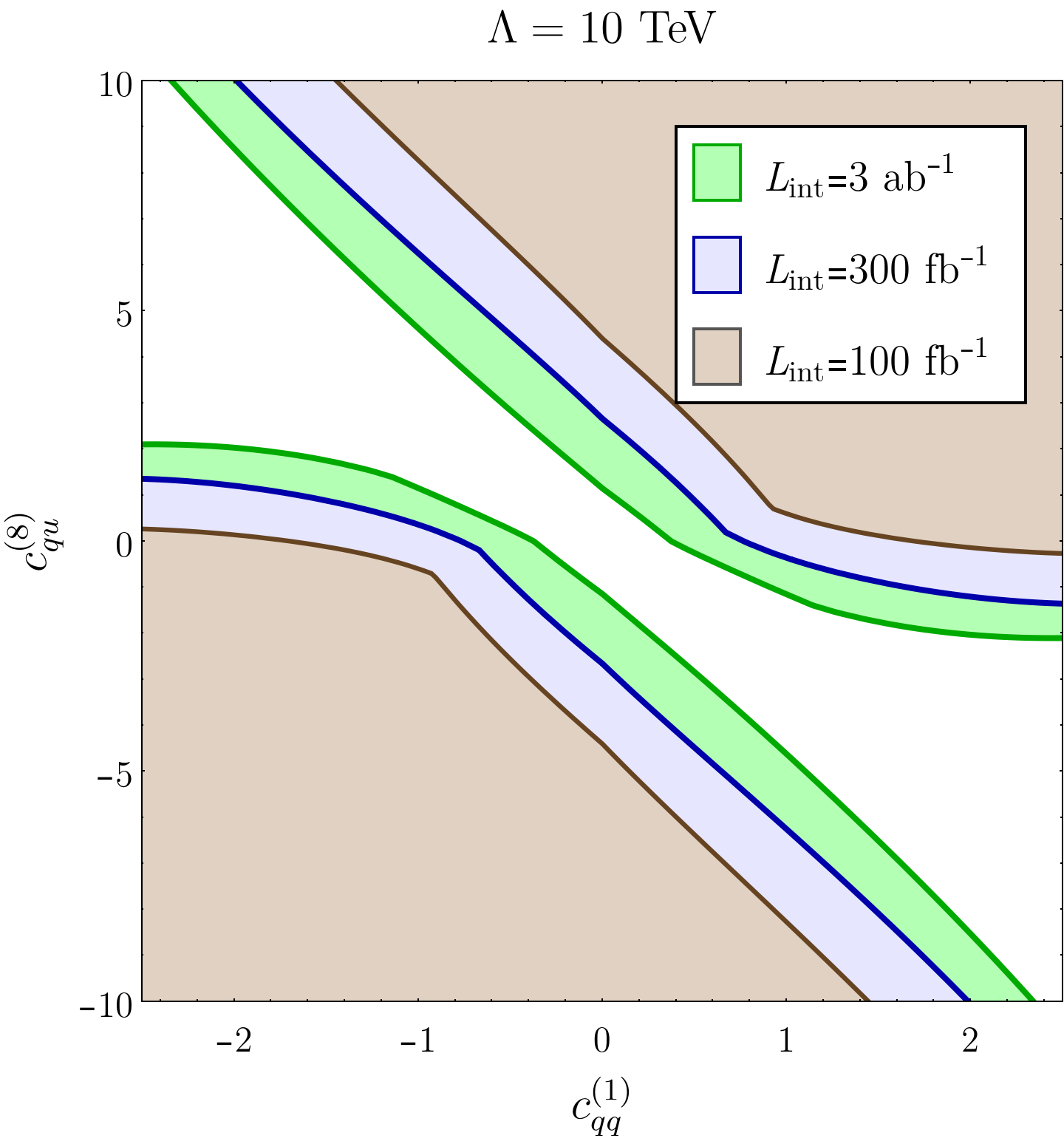}
 \caption{Projected exclusion regions from the unnormalized angular spectrum for the fixed-scale case with $\Lambda=10~\mathrm{TeV}$ at integrated luminosities of 100, 300, and 3000 fb$^{-1}$. The blue regions correspond to $N_8=1$ whereas the orange regions show the exclusions for $N_8=10$. Regions bounded by solid lines show the exclusions when the theory error $\Delta_{\mathrm{theo},1}$ is used, whereas dashed lines illustrate the exclusions when the more conservative error $\Delta_{\mathrm{theo},2}$ is used. Note that, unlike the more usual case of an allowed ellipse, there are two constrained elipses, with a blind direction corresponding to large cancellations between the effects of the two operators. The fourth panel shows the area for small Wilson coefficients in more detail, this time for three different choices of integrated luminosities and $N_8=1$ with theory error model $\Delta_{\text{theo},1}$.}
 \label{fig:FW_theo2}
\end{figure}

The exclusion regions for $\Lambda=10$ TeV and three different integrated luminosities are illustrated in Figure~\ref{fig:FW_theo2}. We consider the luminosities $L_\text{int}=100\text{ fb}^{-1}$, $L_\text{int}=300\text{ fb}^{-1}$ and $L_\text{int}= 3000\text{ fb}^{-1}$, approximating the currently available dataset, the data collected by the end of Run 2 and the data collected by the end of the high-luminosity run. It is important to note that increasing the luminosity does not alter the bounds in the regions where the Wilson coefficients are large. In this regime the analysis is limited by the theory error. Thus, already the current data can provide important information about the constraints on the Wilson coefficients. Another interesting feature is the kinks in the solid contours of the exclusion regions where the theory error defined in eq.~\eqref{eq:theoryError1} is used: this kink arises when the Wilson coefficient is large enough for the function $\Delta_{\text{theo},1}$ to choose the piece independent of the parameter $N_8$. Note that for $N_8=10$ only a relatively small angle of same-sign Wilson coefficients is bounded.

\section{EFT searches in the dijet invariant mass spectrum}
\label{sec:mjj}

When the theory uncertainty from higher-dimensional operators is treated in a more conservative manner, the analyses currently employed to place bounds on contact operators are considerably weakened. We have especially seen that more data is needed in order to obtain constraints depending on which operators are assumed to contribute, most prominently in cases where different operators come with coefficients of opposite signs. In an attempt to mitigate this problem, we explore the possibility of a new analysis, where we jettison the dependence on the angular variable $\chi$ and perform the fit on the distribution of the data across the dijet invariant mass spectrum only. The larger contribution overall arises from $c_{qq}^{(1)}$, so we will neglect $c_{qu}^{(8)}$ and in the theory errors set $c_8=|c_{qq}^{(1)}|$ in the following. Note, however, that the contributions from $c_{qu}^{(8)}$ can be easily reabsorbed into a slightly shifted value of $c_{qq}^{(1)}$, as the angular distinctions are discarded by design here.

For our analysis we employ the $m_{jj}$ binning used in the CMS data and model the systematic uncertainties by summing the ones reported in the CMS analysis over the angular spectrum in each dijet mass bin in quadrature. If we combine the systematic errors CMS reports in the bins in the angular spectrum linearly instead of in quadrature, the bounds weaken considerably, but it is still possible to constrain parameter space. For example, the excluded range of scales in the case $c_{qq}^{(1)}=1$ and $N_8=1$ for an integrated luminosity of $1000~\mathrm{fb}^{-1}$ shrinks to 6~TeV~$<\Lambda<12$~TeV when combining the errors linearly rather than the 5~TeV~$<\Lambda<18$~TeV for combination in quadrature. 

 \begin{figure}
 \centering
 \includegraphics[width=.49\textwidth]{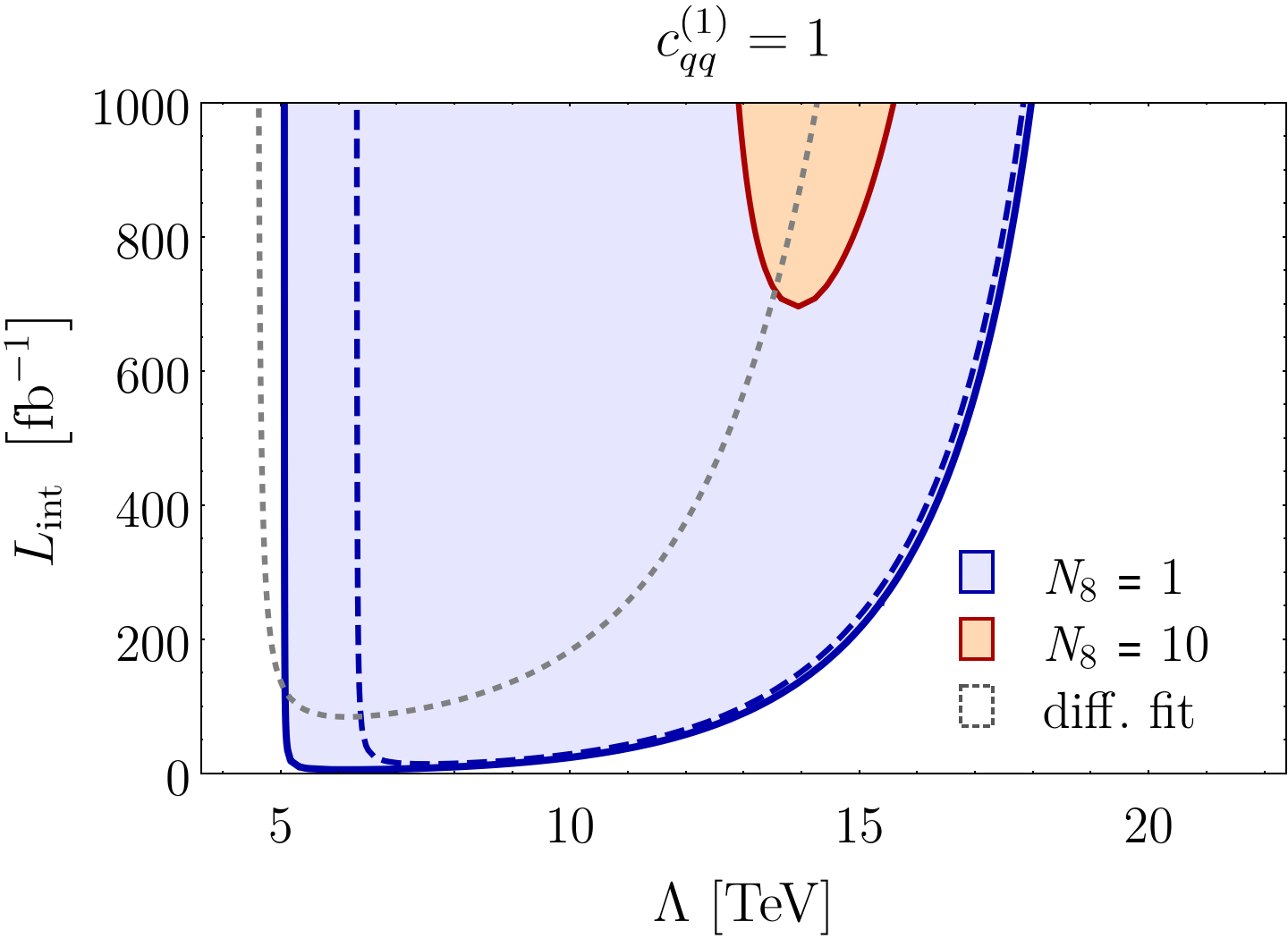}
 \includegraphics[width=.49\textwidth]{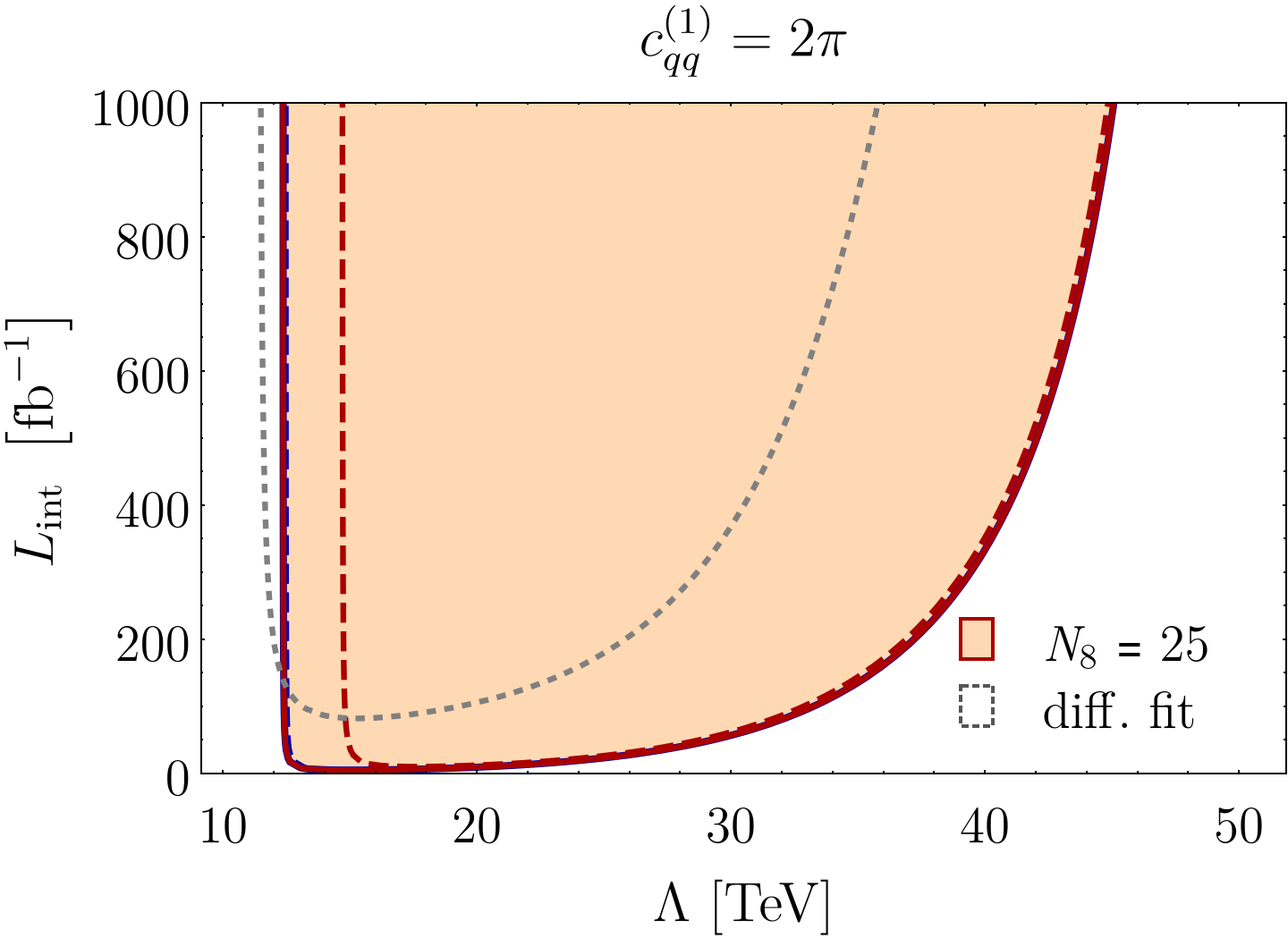}
 \caption{Projected exclusion regions with fixed-Wilson case ($c_{qq}^{(1)}=1$ on the left, $c_{qq}^{(1)}=2\pi$ on the right) at different integrated luminosities, using only the dijet mass spectrum. The blue regions correspond to $N_8=1$ whereas the orange regions show the exclusions for $N_8=10$ in the left panel and $N_8 = 25$ in the right one. Regions bounded by solid lines represent the exclusions for the theory error $\Delta_{\mathrm{theo},1}$, dashed-colored lines illustrate the exclusions when the more conservative error $\Delta_{\mathrm{theo},2}$ is used. The gray-dotted lines outline the exclusion regions from Figure~\ref{fig:FW_theo1_a} with $c_{qu}^{(8)}=0$, $N_8=1$ and theory error $\Delta_{\mathrm{theo},1}$.}
 \label{fig:regions_mjj_c}
\end{figure}
In Figure~\ref{fig:regions_mjj_c} we present our results for the fixed-Wilson-coefficient case with $C_\mathrm{syst}=1$ in the same form as we did when using the angular distributions. As is evident from the gray lines outlining the regions from the previous analysis, the search in the dijet mass spectrum needs less integrated luminosity to exclude parameter space. At the same amount of collected data and equal choices for the Wilson coefficient and the theory error, the search in the dijet mass spectrum excludes NP scales between 1 TeV and 5 TeV higher than in the angular case. It is interesting to note however, that the new lowest constrainable scale is slightly worse than in the previous iteration of the analysis, as can be seen from the gray line extending beyond the left edge of the colored regions. This arises because in the large-statistics regime removing low-$\chi$ events can significantly reduce theoretical uncertainties, though it would also lose distinguishing power between the two relevant combinations of Wilson coefficients; given our lack of knowledge about the full experimental errors, we do not try to construct a fully optimized search for every integrated luminosity.

\begin{figure}
 \centering 
 \includegraphics[width=.49\textwidth]{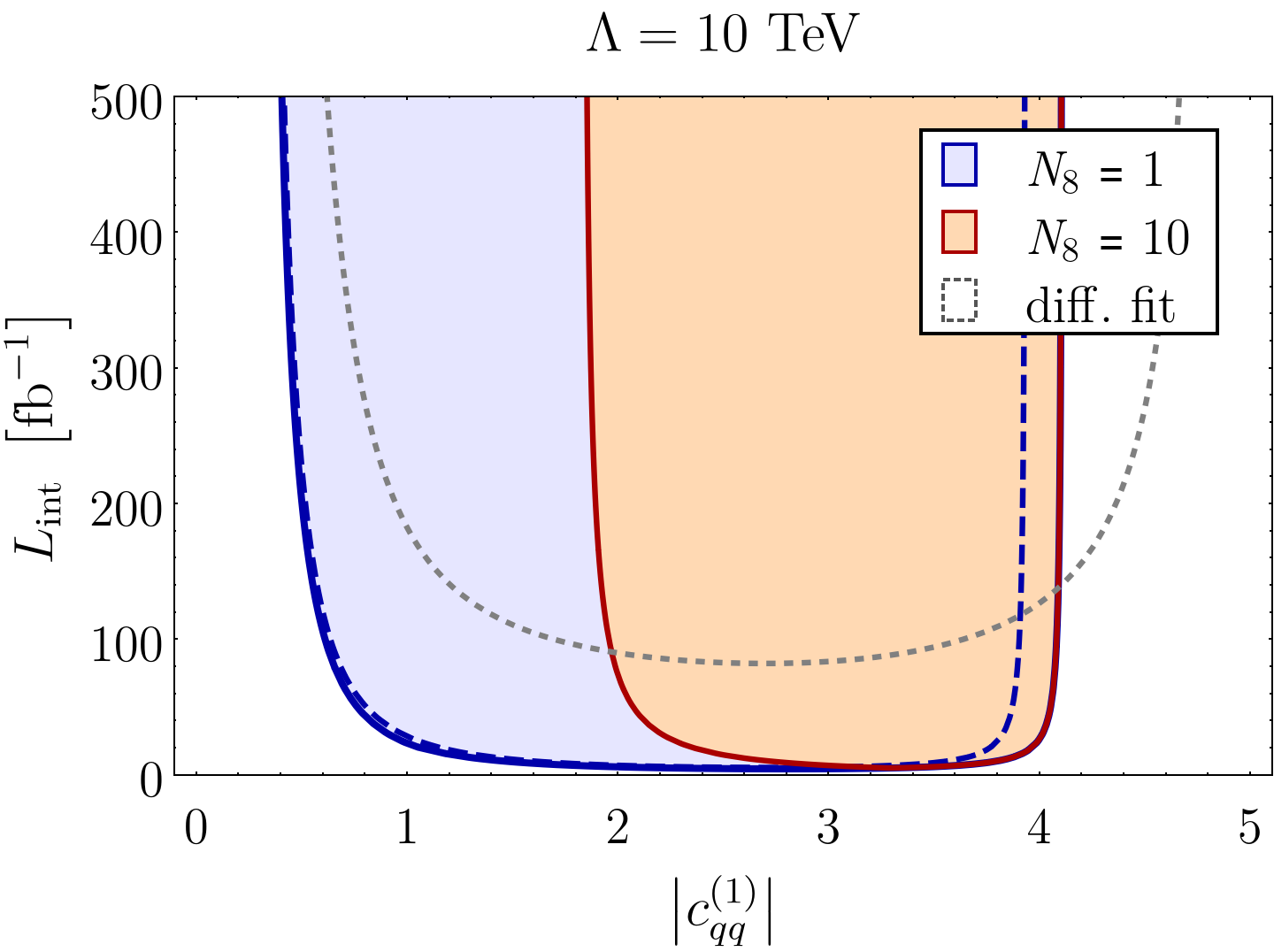}
 \includegraphics[width=.49\textwidth]{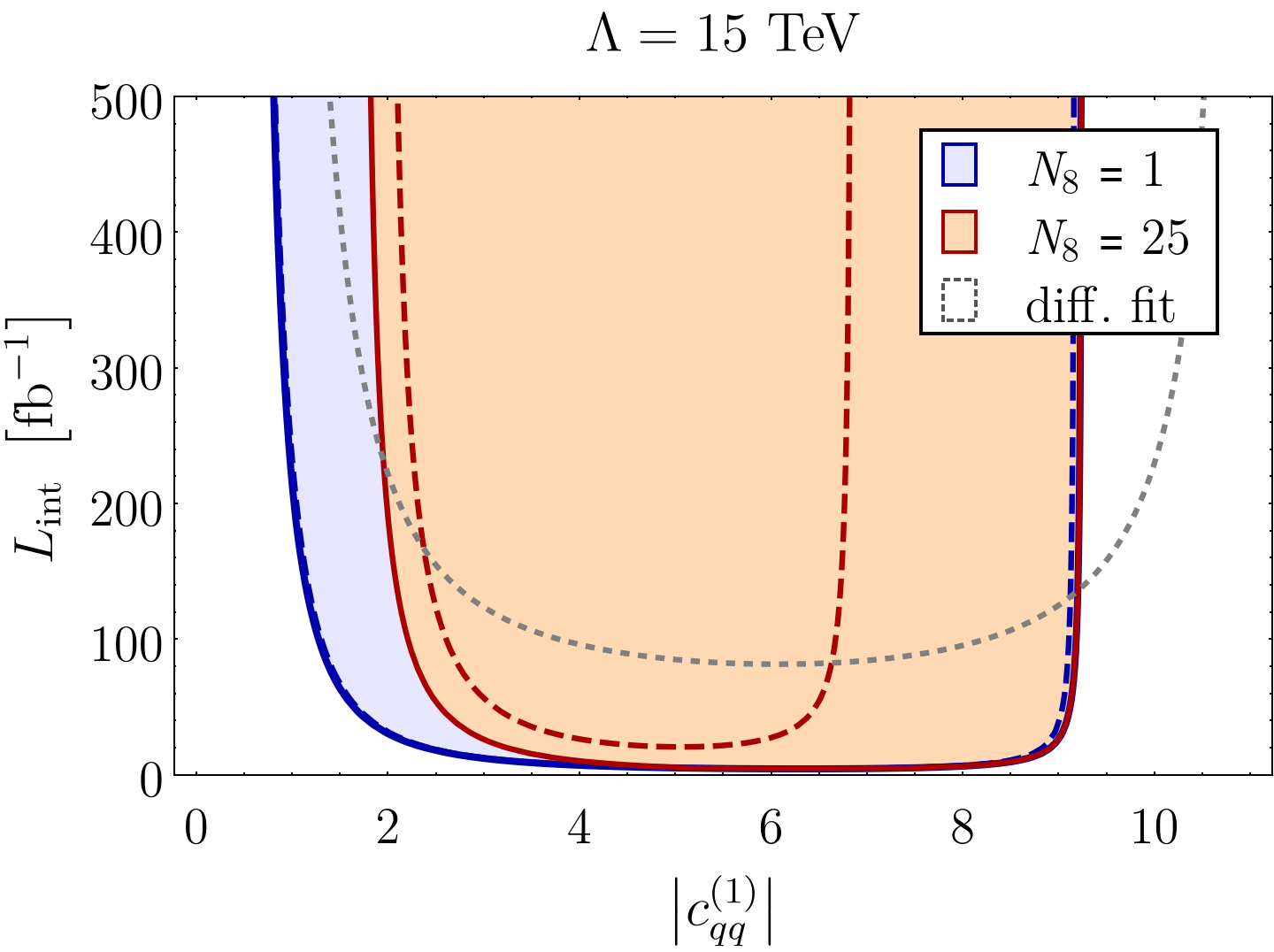}
 \caption{Projected exclusion regions for the fixed-scale case ($\Lambda=10~\mathrm{TeV}$ on the left, $\Lambda=15~\mathrm{TeV}$ on the right) at different integrated luminosities, using only the dijet mass spectrum. The blue regions correspond to $N_8=1$ whereas the orange regions show the exclusions for $N_8=10$ in the left panel and $N_8=25$ in the right one. Regions bounded by solid lines represent the exclusions for the theory error $\Delta_{\mathrm{theo},1}$, dashed-colored lines illustrate the exclusions when the more conservative error $\Delta_{\mathrm{theo},2}$ is used. The gray-dotted lines outline the exclusion regions discussed in Figure~\ref{fig:FW_theo2} when $c_{qu}^{(8)}=0$, $N_8=1$ and theory error $\Delta_{\mathrm{theo},1}$.}
 \label{fig:regions_mjj_Lambda}
\end{figure}

In Figure~\ref{fig:regions_mjj_Lambda}, we show the results for the case where the NP scale is fixed and the Wilson coefficient is varied. Again, we see that exclusions start at lower amounts of collected data. Furthermore, the excluded regions start at lower values for the Wilson coefficients, but extend only to smaller maximum values. This is equivalent to the situation in the fixed-Wilson-coefficient case.

In both cases, we find the results to be quite sensitive to rescaling the systematic uncertainties to cases with $C_\mathrm{syst}>1$. Taking the parameters of the blue region in the left panel of Figure~\ref{fig:regions_mjj_c}, $c_{qq}^{(1)}=1$, $N_8=1$ and $\Delta_{\mathrm{theo},1}$, we find that the excluded region at maximum integrated luminosity shrinks down to $7~\mathrm{TeV}<\Lambda < 11~\mathrm{TeV}$ for $C_\mathrm{syst}=3$ and vanishes completely for $C_\mathrm{syst}=4$.

\section{Conclusions}
\label{sec:conc}

Constraining the SMEFT in a fully UV-independent way allows us to provide bounds on a large class of models of heavy NP, but it requires significant conservatism in search design. As we proceed from the regime of new energy frontiers in which new particles accessible to the collider can quickly reveal themselves and toward an era of constraints based on high-precision measurements, consistent EFT techniques provide the opportunity to encode our measurements in a way which will allow straightforward understanding of the constraints on models of NP, not unlike the pseudo-observables constructed and painstakingly analyzed by the LEP ElectroWeak Working Group.

One of the main lessons from this article, that it is impossible to use events at higher energies than the EFT cutoff scale to constrain EFT effects, was already known at an intuitive level. This is apparent in the fact that all the consistent constraints here do not give lower bounds on the EFT cutoff scale, but rather rule out a range of scales. In fact, the usual scales considered in a SMEFT analysis of Higgs or EW properties of order one to few TeV are totally unconstrained by these searches, no matter the amount of statistics or level of improvement in the systematic errors in future LHC running, simply because the events being used to derive constraints in this analysis are too high energy, and thus it is impossible to derive any meaningful prediction from the EFT. An important avenue of future research will be to revisit dijet production in previous experiments to investigate the extant constraints on such relatively low scales.

The study also emphasizes the importance of understanding as well as possible the systematic errors which contribute to these high-energy searches, particularly those having to do with QCD. The amount of systematic error involved in an unnormalized search for EFT effects in dijets is not known at the moment, but we do see that it must remain comparable to that in the normalized case if meaningful bounds are to be derived. This motivates both further improved understanding of the QCD predictions for this signal and challenging experimental work to fully characterize and minimize the impact of jet energy scale and other uncertainties.

Dijet events present the greatest challenge to consistent interpretation in an EFT, as they generically probe the very highest energies accessible at a hadron collider. Our reanalysis and proposed searches will be able to give the first consistent EFT bounds on four-quark operators from LHC data. These operators have important effects at one-loop order on very precisely measured observables, and as such constraining them at tree level is important to complete our understanding of the constraints on the SMEFT. This article presents a proof of principle that it is possible to derive constraints on EFT effects from LHC data in a fully consistent way, and by having considered the most challenging case indicates that this should be possible to do in a host of other signatures as well. Once a sufficient breadth of observables has been consistently predicted in the SMEFT and searched for in experiments it will become possible to constrain the full set of Wilson coefficients simultaneously, allowing for the straightforward checking of new models against the sum total of precision measurements in particle physics.

\section*{Acknowledgments}
We thank Michael Trott for helpful comments on the manuscript, Klaus Rabbertz from the CMS collaboration for providing us with additional details on the CMS analysis, and Hubert Spiesberger for technical support.
The work of SA and MK was partially supported by the Cluster of Excellence \textit{Precision Physics, Fundamental Interactions and Structure of Matter} (PRISMA-EXC 1098), grant 05H12UME of the German Federal Ministry for Education and Research, and the DFG Graduate School \textit{Symmetry Breaking in Fundamental Interactions} (GRK 1581). The work of WS was supported by the Alexander von Humboldt Foundation, in the framework of the Sofja Kovalevskaja Award 2016, endowed by the German Federal Ministry of Education and Research. SA thanks the Theoretical Physics Department of Fermilab for the hospitality during the research visit that was partially supported by grant ICRADA FRA-2016-0040. MK is also grateful for support by the Swiss National Science Foundation (SNF) under contract 200021-175940. The work of WS was performed in part at the Aspen Center for Physics, which is supported by National Science Foundation grant PHY-1607611.

\bibliography{Sources}

\end{document}